\documentclass[a4paper,fleqn,usenatbib]{mnras}
\usepackage[T1]{fontenc}
\usepackage{ae,aecompl}
\usepackage{graphicx}	
\usepackage{amsmath}	
\usepackage{amssymb}	
\usepackage{booktabs}
\usepackage{times}
\usepackage{soul}
\usepackage{caption,xcolor}
\usepackage{subcaption}
\usepackage{threeparttable}
\usepackage{longtable}

\title[Polarization towards NGC 2345]{Polarization, Polarizing Efficiency, and  Grain alignment towards the direction of the cluster NGC 2345}
\author[Singh et al.]{
Sadhana Singh$^{1,2}$\thanks{E-mail: ssingh@aries.res.in}, 
Jeewan C. Pandey$^{1}$ \thanks{jeewan@aries.res.in}, 
Thiem Hoang$^{3,4}$\thanks{thiemhoang@kasi.re.kr}
\\
$^{1}$Aryabhatta Research Institute of Observational Sciences (ARIES), Manora Peak, Nainital 263001, India\\
$^{2}$School of Studies in Physics \& Astrophysics, Pt. Ravishankar Shukla University, Raipur-492010, India\\
$^3$ Korea Astronomy and Space Science Institute, Daejeon 34055, Republic of Korea\\
$^4$ Korea University of Science and Technology, 217 Gajeong-ro, Yuseong-gu, Daejeon, 34113, Republic of Korea\\
}

\date{Accepted ... Received ; in original form }

\pubyear{-}

\begin{document}
\label{firstpage}
\pagerange{\pageref{firstpage}--\pageref{lastpage}}
\maketitle
\begin{abstract}
We have investigated the grain alignment and dust properties towards the direction of the cluster NGC 2345 using the multi-band optical polarimetric observations. For the majority of the stars, the observed polarization is found to be due to the interstellar medium with average values of maximum polarization and wavelength corresponding to it as 1.55\% and 0.58 $\mu m$, respectively. This reveals a similar size distribution of dust grains to that in the general interstellar medium in the direction of NGC 2345. Alteration of dust properties near the distance of 1.2 kpc towards the direction of NGC 2345 has been noticed. The dust grains located beyond this distance are found to be aligned with the Galactic magnetic field, whereas a dispersion in orientation of the dust grains lying in the foreground of this distance is found. Polarizing efficiency of grains in this direction is found to be close to the average efficiency for our Galaxy. The decreased grain size along with the increased polarizing efficiency towards the core region of the cluster indicates the local radiation field is higher within the cluster which is responsible for the increased alignment efficiency of small grains. The wavelength of maximum polarization (associated with the average size of aligned grains) is also found to increase with extinction and reduces with the increase in polarizing efficiency, which can be explained by the radiative torque alignment mechanism. 

\end{abstract}


\section{Introduction}\label{intro}
The starlight passing through the interstellar medium (ISM) becomes linearly polarized due to the differential extinction from the alignment of asymmetric dust grains in the ISM \citep{1949Sci...109..166H,1949Sci...109..165H}. The alignment process of dust grains has been a topic of discussion in the past several decades \citep[see e.g.][]{1951ApJ...114..206D,1967ApJ...147..943J,1978ppim.book.....S,1983ApJ...272..551A,1986ApJ...308..281M,2013lcdu.confE...3V}. The paramagnetic relaxation and radiative alignment torque are the main mechanisms considered for the alignment of dust grains.  In the paramagnetic relaxation mechanism, initially the grains spin after collision with hydrogen atoms and ions, and the spin axis of grains aligns with the local magnetic field by dissipation of the magnetization energy \citep{1951ApJ...114..206D}. However, in order to align grains on time-scales faster than disruption, there are further advancements in this theory by including superparamagnetic grains or superthermal radiation etc. \citep{1967ApJ...147..943J,1979ApJ...231..404P,1986ApJ...308..281M}.

The radiative torque alignment is considered to be the best alignment mechanism to explain the optical and infrared polarization \citep{1976Ap&SS..43..291D,1996ApJ...470..551D,2007MNRAS.378..910L}. Radiative torque alignment is due to the interaction of anisotropic radiation (wavelength of radiation less than the diameter of grains) with an irregular shaped grain. Dust grains are aligned in such a way that the short axis of grains is parallel to the magnetic field. According to this theory, for a strong and anisotropic radiation field, the alignment direction can be different from the magnetic field and it may be towards the direction of radiation \citep{2007MNRAS.378..910L,2021ApJ...908...12L}. In the majority of environments, the radiative torque alignment is a magnetic alignment in which grains are aligned along the magnetic field and the plane of the sky projection of the field can be tracked by the direction of observed polarization due to the dichroic extinction \citep{2016ApJ...831..159H}. Presently, all these mechanisms are being explored to support most observational findings. Thus, numerous observational results towards different directions of Galaxy are very important and play a very effective role in better understanding of the grain alignment. The polarimetric study will be an effective tool for providing information about the magnetic field orientation \citep{1996ApJ...462..316H,2003JQSRT..79..881L,2007JQSRT.106..225L}, also it reveal information about the properties of ISM dust grains and their alignment \citep[e.g.][]{1992ASNYN...4a..15W,1994ApJ...431..783K,2012JQSRT.113.2334V,2021AJ....161..149S}. Dust polarization is associated with the fraction of aligned dust grains present in the ISM and how efficient the alignment will be, it is also related to the particular size and type (composition) of grains. It also depends on the magnetic field geometry along the line of sight. The correlation of polarization degree with reddening was reported by  \citet{1949ApJ...109..471H,1956ApJS....2..389H} with the existence of an upper limit on polarization, and in ongoing observational studies it was seen for the majority of lines of sight \citep[e.g.][etc.]{2020AJ....160..256S,2020AJ....159...99S,2010MNRAS.403.1577M,2007A&A...462..621V}. The polarimetric study in different bands leads to information about the size of grains. The variation of polarization with wavelength is generally followed by Serkowski relation \citep{1975ApJ...196..261S} for ISM polarization and is associated with parameters like maximum polarization ($P_{\rm max}$), and wavelength ($\lambda_{\rm max}$) corresponding to $P_{\rm max}$, which is a measure of size distribution of dust grains \citep[e.g.][]{1994ApJ...431..783K}. Additionally, details of the change in the interstellar environment and magnetic fields, the existence of any dust layer, the concentration of foreground dust etc. can be acquired.

In this paper, we present a multi-band polarimetric study for an open star cluster NGC 2345. The open cluster NGC 2345 is a young cluster in Canis Major constellation ($l$ = 226\degr.58, $b$ = -2\degr.31). Distance, reddening [$E(B-V)$], and age of this cluster were estimated to be in the range of 2.2-3.0 kpc, 0.59-0.68 mag, and 55-79 Myr \citep[in][]{2005A&A...438.1163K,2013A&A...558A..53K,2015AJ....149...12C,2018A&A...618A..93C,2019A&A...631A.124A}. \cite{2005A&A...438.1163K} have derived the angular radius of core and cluster as 4\arcmin.2 and 7\arcmin.2, respectively, whereas \citet{2019A&A...631A.124A} have estimated core radius of 3.44$\pm$0.08 arcmin and tidal radius of 18.7$\pm$1.2 arcmin for the cluster. Non radial distribution of dust associated with the cluster was noticed in various  past studies with a variable reddening $E(B-V)$ from 0.4 to 1.2 mag \citep[like][]{1974A&AS...16...33M,2015AJ....149...12C,2019A&A...631A.124A}. \citet{2015AJ....149...12C} have found variable extinction across the cluster and an overdensity was noticed upto a radius of $\approx$ 3.75 arcmin, which is a cluster radius. 

The paper is organized as follows: the observations and data reduction are given in Section \ref{obs}, the results and analysis are described in Section \ref{res}, whereas the discussion and summary are described in Section \ref{diss} and \ref{summ}, respectively. 

\section{Observations and data reduction}\label{obs}
Polarimetric observations of the cluster NGC 2345 and a field region (which is $\sim$ 1\degr\, away from the centre of NGC 2345, to increase the field population to improve statistics) were carried out on 22 January, and 14 and 15 February 2018 using ARIES imaging polarimeter \citep[AIMPOL;][]{2004BASI...32..159R} which is mounted as a back-end instrument at the 104-cm telescope (f/13 Cassegrain focus) of ARIES. The AIMPOL consist of a half-wave plate (HWP) and a Wollaston prism. HWP is rotatable and we performed observations at four different positions of HWP i.e. 0\degr, 22\degr.5, 45\degr, 67\degr.5 from the celestial north-south direction. Through the Wollaston prism, each image is transformed into ordinary (o-) and extraordinary (e-) images on the detector. The detector used was a 1k$\times$1k charge-coupled camera (CCD) cooled by liquid nitrogen. The read noise and gain of CCD are 7.0 $e^{-}$ and 11.98 $e^{-}/ADU$, respectively. The field of view of AIMPOL is $\sim$ 8\arcmin\, in diameter. The observations were performed in four pass-bands $B$ ($\lambda_{eff}$ = 0.44 $\mu m$), $V$ ($\lambda_{eff}$ = 0.55 $\mu m$), $R$ ($\lambda_{eff}$ = 0.67 $\mu m$),  and $I$ ($\lambda_{eff}$ = 0.80 $\mu m$). Exposure times were 250, 200, 120, and 150 seconds in $B$, $V$, $R$, and $I$ bands, respectively. We have taken at least three frames at each position of HWP. All frames at a position are summed in order to get a good signal to noise ratio. Aperture photometry was performed to get the fluxes of o- and e- images using Image Reduction and Analysis Facility\footnote{iraf.noao.edu}. Further reduction techniques are given in \citet{2020AJ....159...99S}. Correction for instrumental polarization was performed using an unpolarized standard star (HD21447). The instrumental polarization was found to be $\leq$ 0.3\% in all bands. We have observed two standard polarized stars (HD19820 and HD25443) for zero point polarization angle correction. The observed values for polarized standard stars are found to be in good agreement with the standard values as given in \citet{1992AJ....104.1563S}.

As there is no grid in AIMPOL, the overlapping of images can not be avoided. Thus we have manually selected the isolated sources in each $B$, $V$, $R$, and $I$ bands. Owing to this and considering the offset of the observed regions in different days, we have selected 172 stars in $B$, 146 in $V$, 168 in $R$, and 156 in $I$ bands. Further, we have discarded those values of a polarization for which the relative error is more than  0.5  (or $>2 \sigma$).  Adopting this additional criteria, we are left with 151, 132, 157, 144 stars in  $B$, $V$, $R$, and $I$ bands, respectively.
A total of 197 sources, of which 135 are of the cluster region and 62 from the field region are included in this study. The astrometry was performed with the online available tool\footnote{https://nova.astrometry.net/upload}. For the identification of observed stars, we have cross-matched our stars with the \textit{Gaia} DR2 data sets \citep{2016A&A...595A...1G, 2018A&A...616A...1G}. Most of the stars were matched with the \textit{Gaia} sources within 2\arcsec\, offset, only 3 stars were matched between 2\arcsec\ to 3.8\arcsec\, offset. All these sources are fulfilled by the condition for a good astrometric solution as given in \citet[][]{2018A&A...616A...2L}. 
Membership information of observed stars in the cluster NGC 2345 are taken from \citet{2018A&A...618A..93C}. We have considered stars as members of the cluster if the membership probability (MP) as given in \citet{2018A&A...618A..93C} is more than 50\%. 

\section{Results and Analysis}\label{res}
The degree of polarization ($P$) and polarization position angle ($\theta$) in four bands $B$, $V$, $R$, and $I$ for observed sources are given in Table \ref{tab:pthe}. The source ID designated in our work is in the first column and the second column is the identification ID taken from \textit{Gaia}. The offset from \textit{Gaia} position is given in column 3. Then consecutive columns represent $P$ and $\theta$ with associated errors in $B$, $V$, $R$, and $I$ bands. In this table, the first 135 sources are from the cluster region and the rest 62 sources are from the field region. The member stars of the cluster NGC 2345 are marked with an asterisk symbol with their IDs.   

\onecolumn
{\fontsize{6.6}{7.8}\selectfont
\begin{longtable}{llccccccccr}
\caption{The degree of polarization ($P$) and position angle ($\theta$) in $B$, $V$, $R$, and $I$ bands for all observed stars in the cluster NGC 2345 and a field region.}
\label{tab:pthe}\\
\hline
ID & \textit{Gaia} ID & Offset(\arcsec)& $P_{B}\left(\%\right)$ & $\theta_{B}\left({^o}\right)$ & $P_{V}\left(\%\right)$ & $\theta_{V}\left({^o}\right)$ & $P_{R}\left(\%\right)$ & $\theta_{R}\left({^o}\right)$ & $P_{I}\left(\%\right)$ & $\theta_{I}\left({^o}\right)$\\
\hline
&&&&&Cluster region \\
\hline
1*       &  3044669369450492288 &  0.46    & 	1.16$\pm$0.12  &  162.0$\pm$2.8	 &  		-  & -  		        &    	1.79$\pm$0.12  &  148.9$\pm$1.9     &     1.84$\pm$0.23  &  160.3$\pm$3.6  \\    
2       &  3044669369450499328 &  0.39    &    	0.92$\pm$0.42  &  119.6$\pm$13.4	 &  		- & -   		        &    	0.29$\pm$0.11  &  91.4 $\pm$10.4    &   	  -      &     -   \\  
3*       &  3044666380153268992 &  0.58    &    	0.97$\pm$0.45  &  154.8$\pm$13.2	 &  		-	 & -  	        &    	1.01$\pm$0.13  &  145.8$\pm$3.6     &     1.66$\pm$0.02  &  155.4$\pm$0.4  \\   	
4*       &  3044665486800115200 &  0.61    &    	1.63$\pm$0.19  &  155.4$\pm$3.2	 &  		-	  	&-        &    	1.51$\pm$0.14  &  154.8$\pm$2.6     &     1.68$\pm$0.38  &  151.7$\pm$6.5  \\   	
5       &  3044666204049384704 &  0.52    &    	1.22$\pm$0.24  &  162.8$\pm$5.5	 &  		-	   	  &-      &    	1.78$\pm$0.17  &  145.0$\pm$2.8     &     1.71$\pm$0.06  &  158.8$\pm$1.1  \\   	
6*       &  3044668682255788672 &  0.40    &    	2.89$\pm$0.13  &  139.3$\pm$1.3	 &     2.62$\pm$0.01   &  129.4$\pm$0.1     &    	3.13$\pm$0.14  &  137.7$\pm$1.3     &     2.73$\pm$0.03  &  138.8$\pm$0.4  \\   
7*       &  3044665727318289152 &  0.29    &    	1.57$\pm$0.08  &  145.5$\pm$1.5	 &  	-        &       -        &    	3.15$\pm$0.18  &  139.9$\pm$1.7     &    	-       &       -     \\   	
8*       &  3044668819694724352 &  0.24    &    	4.48$\pm$1.64  &  163.9$\pm$10.4	 &	-   &      -        &    	1.47$\pm$0.12  &  148.2$\pm$2.2     &     1.40$\pm$0.54  &  165.2$\pm$11.0 \\   		
9*       &  3044669575608944256 &  0.21    &    	1.75$\pm$0.02  &  153.1$\pm$0.4	 &  	-        &       -        &    	1.75$\pm$0.13  &  146.4$\pm$2.2     &     1.65$\pm$0.11  &  146.9$\pm$2.0  \\   		
10      &  3044668888404742528 &  0.26    &    	1.97$\pm$0.04  &  152.9$\pm$0.6	 &  	-        &       -        &    	1.59$\pm$0.01  &  150.8$\pm$0.2     &     1.54$\pm$0.15  &  158.2$\pm$2.8  \\ 
11*      &  3044669541249212800 &  0.11    &    	2.86$\pm$0.18  &  129.5$\pm$1.8	 &  	-        &       -        &    	2.12$\pm$0.01  &  127.9$\pm$0.1     &     1.78$\pm$0.29  &  129.3$\pm$4.7  \\   	
12*      &  3044669094572631040 &  0.11    &    	2.23$\pm$0.04  &  157.9$\pm$0.5	 &  	-        &        -       &    	1.07$\pm$0.19  &  152.0$\pm$5.0     &     1.49$\pm$0.30  &  161.3$\pm$5.7  \\   	
13*      &  3044669163292102144 &  0.06    &    	1.33$\pm$0.01  &  154.2$\pm$0.2	 &  	-        &       -        &    	1.53$\pm$0.14  &  150.1$\pm$2.5     &     1.49$\pm$0.15  &  153.1$\pm$2.9  \\   		
14*      &  3044664215489824256 &  0.84    &    	0.95$\pm$0.17  &  113.5$\pm$5.2	 &     1.01$\pm$0.06   &  133.2$\pm$1.8     &    	1.21$\pm$0.09  &  120.3$\pm$2.2     &     0.89$\pm$0.18  &  134.4$\pm$6.0  \\   
15*      &  3044668922773965824 &  0.48    &    	1.75$\pm$0.08  &  164.7$\pm$1.3	 &  	-        &       -        &    	2.29$\pm$0.04  &  155.4$\pm$0.4     &   	   -     &       -     \\   		
16      &  3044665761678349056 &  0.19    &    	0.99$\pm$0.08  &  11.0 $\pm$2.4	 &  	-     &      -        &    	0.72$\pm$0.04  &  50.9 $\pm$1.7     &     1.33$\pm$0.25  &  27.1 $\pm$5.5  \\   		
17*      &  3044669644328400896 &  0.83    &    	0.78$\pm$0.01  &  101.2$\pm$0.5	 &     1.12$\pm$0.04   &  138.5$\pm$1.0     &    	1.86$\pm$0.06  &  139.7$\pm$0.9     &     1.72$\pm$0.62  &  135.6$\pm$10.4 \\   
18*      &  3044669472529743360 &  1.18    &    	2.49$\pm$0.17  &  154.8$\pm$1.9	 &  	-        &               &    	1.67$\pm$0.22  &  146.6$\pm$3.9     &     0.94$\pm$0.03  &  139.0$\pm$0.9  \\   		
19*      &  3044666414512997248 &  0.55    &    	    -      &       -	 &  	-        &              - &    	1.35$\pm$0.07  &  150.9$\pm$1.5     &     1.21$\pm$0.06  &  164.6$\pm$1.3  \\       		   
20*      &  3044666307128599040 &  0.59    &    	    -    &       -	 &  	-        &              - &    	1.62$\pm$0.04  &  141.7$\pm$0.7     &     1.41$\pm$0.15  &  145.2$\pm$3.1  \\       		   
21      &  3044665967836424192 &  0.55    &    	    -      &  -    	 &  	  -      &              - &    	1.00$\pm$0.12  &  140.6$\pm$3.4     &     1.33$\pm$0.12  &  159.1$\pm$2.5  \\       
22*      &  3044669266371312384 &  0.69    &    	    -      &  -   	 &  	  -      &  -               &    	2.47$\pm$0.03  &  164.8$\pm$0.4     &     2.84$\pm$0.22  &  168.5$\pm$2.3  \\       		   
23*      &  3044669266371307776 &  0.70    &    	    -     &  -    	 &  	  -      &  -                &    	2.23$\pm$0.25  &  170.3$\pm$3.2     &     2.38$\pm$0.21  &  168.3$\pm$2.5  \\       		  
24      &  3044675622922858240 &  0.26    &    	1.29$\pm$0.24  &  157.2$\pm$5.2	 &     1.82$\pm$0.04   &  154.7$\pm$0.6     &       1.63$\pm$0.09  &  154.4$\pm$1.5     &     1.59$\pm$0.04  &  158.7$\pm$0.8  \\ 	       	
25*      &  3044675588563126656 &  0.41    &    	1.59$\pm$0.06  &  131.9$\pm$1.1      &     1.88$\pm$0.11   &  127.9$\pm$1.7     &    	2.09$\pm$0.03  &  124.6$\pm$0.4     &     1.60$\pm$0.10  &  161.0$\pm$1.8  \\       	
26      &  3044675588563129088 &  0.39    &    	1.62$\pm$0.11  &  12.1 $\pm$1.9      &     1.23$\pm$0.17   &  109.7$\pm$4.1     &    	1.13$\pm$0.01  &  145.1$\pm$0.3     &     0.53$\pm$0.07  &  177.6$\pm$3.7  \\       		
27      &  3044677237830539136 &  0.45    &    	1.04$\pm$0.33  &  140.6$\pm$9.0      &     0.76$\pm$0.18   &  142.1$\pm$6.8     &    	1.05$\pm$0.02  &  156.2$\pm$0.5     &     0.60$\pm$0.05  &  136.7$\pm$2.2  \\       		
28*      &  3044676550635786880 &  0.22    &    	1.04$\pm$0.01  &  163.0$\pm$0.3      &     1.29$\pm$0.05   &  152.0$\pm$1.2     &    	1.42$\pm$0.22  &  149.8$\pm$4.4     &     0.64$\pm$0.12  &  146.1$\pm$5.3  \\       
29*      &  3044676378837101056 &  0.25    &    	2.66$\pm$0.41  &  153.8$\pm$4.4      &     1.41$\pm$0.35   &  147.2$\pm$7.1     &    	2.07$\pm$0.02  &  143.5$\pm$0.3     &     1.82$\pm$0.29  &  149.3$\pm$4.6  \\       		
30      &  3044864532763157248 &  0.30    &    	0.61$\pm$0.07  &  167.7$\pm$3.3      &     0.36$\pm$0.04   &  30.9 $\pm$3.0     &    	0.65$\pm$0.10  &  3.1  $\pm$4.5     &     1.22$\pm$0.08  &  7.5  $\pm$1.8  \\       		
31      &  3044864395324215168 &  0.17    &    	1.62$\pm$0.18  &  124.0$\pm$3.2      &     0.90$\pm$0.18   &  165.9$\pm$5.8     &    	-        &        -       &   	   -      &  -     \\       		
32*      &  3044676688074753536 &  0.11    &    	    -     &       - 	 &     1.26$\pm$0.18   &  11.3 $\pm$4.0     &       0.91$\pm$0.23  &  176.5$\pm$7.1     &     1.12$\pm$0.04  &  179.6$\pm$1.1  \\ 		
33      &  3044670503321866752 &  0.58    &    	1.12$\pm$0.04  &  39.7 $\pm$1.0      &     0.62$\pm$0.04   &  51.5 $\pm$1.8     &    	0.35$\pm$0.02  &  151.3$\pm$1.8     &     0.66$\pm$0.01  &  13.2 $\pm$0.1  \\       		
34*      &  3044670572041356160 &  0.31    &    	1.38$\pm$0.18  &  176.1$\pm$3.6      &     3.33$\pm$0.10   &  159.2$\pm$0.9     &    	2.25$\pm$0.03  &  141.5$\pm$0.4     &     2.08$\pm$0.01  &  149.6$\pm$0.1  \\       		
35      &  3044864734622541696 &  0.23    &    	1.35$\pm$0.08  &  43.7 $\pm$1.7      &     1.58$\pm$0.04   &  35.5 $\pm$0.6     &    	1.02$\pm$0.24  &  30.9 $\pm$6.7     &     1.51$\pm$0.23  &  24.9 $\pm$4.3  \\       		
36      &  3044670812559502464 &  0.13    &    	    -     &       -	 &     1.71$\pm$0.10   &  179.5$\pm$1.7     &       1.65$\pm$0.11  &  165.0$\pm$1.9     &     1.34$\pm$0.25  &  162.5$\pm$5.4  \\	      	   
37*      &  3044670675120561408 &  0.18    &    	1.09$\pm$0.09  &  165.6$\pm$2.3	 &     0.95$\pm$0.30   &  170.1$\pm$9.1     &       2.13$\pm$0.47  &  164.9$\pm$6.3     &     1.67$\pm$0.24  &  152.5$\pm$4.2  \\	       	
38*      &  3044670572041366400 &  0.84    &    	1.29$\pm$0.25  &  138.4$\pm$5.6	 &     2.34$\pm$0.61   &  176.8$\pm$7.4     &       1.62$\pm$0.24  &  140.7$\pm$4.3     &   	    - &  -       \\	       	
39*      &  3044864360964480896 &  0.07    &    	0.75$\pm$0.17  &  17.2 $\pm$6.4	 &     1.07$\pm$0.03   &  177.1$\pm$0.9     &       1.02$\pm$0.07  &  38.0 $\pm$2.1     &   	0.97$\pm$0.37  &  178.2$\pm$10.8 \\	       	
40*      &  3044670606401093504 &  0.43    &    	1.47$\pm$0.09  &  97.7 $\pm$1.7	 &   	 -      &    -           &     	2.62$\pm$0.07  &  166.7$\pm$0.8     &   	1.38$\pm$0.28  &  10.2 $\pm$5.7  \\       		   
41*      &  3044676756794624256 &  0.22    &    	0.71$\pm$0.27  &  175.1$\pm$10.9	 &     0.84$\pm$0.21   &  165.6$\pm$7.1     &       0.79$\pm$0.24  &  174.0$\pm$8.8     &   	1.90$\pm$0.52  &  15.9 $\pm$7.8  \\	       	
42*      &  3044676447556579712 &  0.17    &    	1.30$\pm$0.39  &  143.8$\pm$8.5  	 &     1.34$\pm$0.09   &  166.8$\pm$1.9     &       1.31$\pm$0.27  &  167.0$\pm$5.9     &   	1.37$\pm$0.04  &  165.7$\pm$0.9  \\	       	
43*      &  3044676340174612608 &  0.25    &    	    -     &       -	 &     1.37$\pm$0.18   &  157.2$\pm$3.8     &       1.01$\pm$0.03  &  152.0$\pm$0.8     &   	1.42$\pm$0.11  &  165.4$\pm$2.3  \\ 		   
44      &  3044669713047878656 &  0.98    &    	1.00$\pm$0.05  &  137.5$\pm$1.4      &     1.20$\pm$0.18   &  145.0$\pm$4.2     &    	1.47$\pm$0.09  &  129.6$\pm$1.7     &   	1.69$\pm$0.12  &  123.3$\pm$2.0  \\       		   
45      &  3044864601482633088 &  0.57    &    	    -     &       -	 &  	   -     & -                &       1.31$\pm$0.24  &  177.1$\pm$5.2     &   	1.02$\pm$0.01  &  148.7$\pm$0.1  \\ 		   
46*      &  3044864842000808192 &  0.52    &    	1.63$\pm$0.02  &  158.9$\pm$0.4	 &     1.76$\pm$0.12   &  166.5$\pm$1.9     &       1.48$\pm$0.11  &  158.9$\pm$2.1     &   	2.06$\pm$0.18  &  156.9$\pm$2.6  \\	       	
47      &  3044670537681591936 &  1.16    &    	0.73$\pm$0.18  &  51.2 $\pm$7.1	 &     0.35$\pm$0.05   &  41.2 $\pm$4.3     &       0.33$\pm$0.04  &  55.2 $\pm$3.7     &   	0.21$\pm$0.08  &  179.9$\pm$10.6 \\	       	
48*      &  3044670503321865728 &  0.42    &    	3.33$\pm$0.77  &  69.1 $\pm$6.5      &     2.54$\pm$0.03   &  133.3$\pm$0.3     &    	1.58$\pm$0.35  &  122.4$\pm$6.4     &   	2.65$\pm$0.62  &  124.3$\pm$6.7  \\       		  
49*      &  3044670400242658176 &  0.60    &    	    -     &       -	 &     4.81$\pm$0.70   &  141.5$\pm$4.1     &       3.62$\pm$0.04  &  145.2$\pm$0.3     &   	3.38$\pm$0.31  &  136.5$\pm$2.6  \\ 		  
50*      &  3044670464659348224 &  0.36    &    	1.16$\pm$0.25  &  140.7$\pm$6.3      &     0.53$\pm$0.01   &  145.4$\pm$0.7     &    	1.18$\pm$0.36  &  146.1$\pm$8.9     &   	0.87$\pm$0.40  &  155.7$\pm$13.0 \\       		   
51*      &  3044856591364735232 &  0.51    &    	1.84$\pm$0.11  &  124.5$\pm$1.7      &     1.12$\pm$0.11   &  123.6$\pm$2.9     &    	1.66$\pm$0.10  &  123.5$\pm$1.7     &   	0.82$\pm$0.30  &  121.0$\pm$10.6 \\       		   
52      &  3044856458224709632 &  0.28    &    	1.52$\pm$0.56  &  25.6 $\pm$10.6     &     1.23$\pm$0.11   &  28.9 $\pm$2.5     &       1.15$\pm$0.36  &  50.3 $\pm$9.0     &   	0.41$\pm$0.10  &  179.1$\pm$6.6  \\         	
53      &  3044856561303914496 &  0.43    &    	    -     &       -         &     0.76$\pm$0.13   &  10.2 $\pm$5.1     &       0.45$\pm$0.17  &  39.7 $\pm$10.7    &   	0.45$\pm$0.01  &  44.8 $\pm$0.1  \\ 		  
54      &  3044856183346803584 &  0.15    &    	0.55$\pm$0.07  &  22.1 $\pm$3.6      &     0.58$\pm$0.01   &  14.8 $\pm$0.5     &       0.21$\pm$0.05  &  21.1 $\pm$7.1     &   -     &  -        \\         	
55      &  3044855771029950592 &  0.55    &    	    -     &      -	 &     0.77$\pm$0.07   &  53.0 $\pm$2.5     &       0.12$\pm$0.02  &  10.9 $\pm$3.6     &   	0.82$\pm$0.07  &  152.2$\pm$2.5  \\ 		  
56      &  3044855594932139264 &  0.38    &    	1.26$\pm$0.18  &  5.5  $\pm$4.2  	 &     0.39$\pm$0.17   &  171.7$\pm$12.2    &    	0.83$\pm$0.12  &  151.6$\pm$4.2     &   	0.54$\pm$0.12  &  174.2$\pm$6.3  \\	       	
57      &  3044855736670209536 &  0.41    &    	1.08$\pm$0.03  &  15.6 $\pm$0.9	 &     1.07$\pm$0.19   &  32.8 $\pm$5.2     &       0.23$\pm$0.04  &  65.7 $\pm$4.7     &   	1.05$\pm$0.23  &  17.9 $\pm$6.4  \\	       
58*      &  3044668059477483264 &  0.31    &    	1.43$\pm$0.33  &  134.8$\pm$6.7	 &     0.33$\pm$0.13   &  152.3$\pm$11.3    &       0.53$\pm$0.14  &  110.2$\pm$7.7     &   	0.87$\pm$0.20  &  31.7 $\pm$6.6  \\	     
59      &  3044858141851970944 &  2.42    &    	1.38$\pm$0.27  &  18.4 $\pm$5.6	 &   	  -      &     -         &     	0.90$\pm$0.10  &  5.7  $\pm$3.2     &   	1.51$\pm$0.41  &  173.4$\pm$7.8  \\       	
60      &  3044856355145484160 &  0.29    &    	2.57$\pm$0.41  &  151.6$\pm$4.5	 &     1.37$\pm$0.27   &  176.4$\pm$5.6     &       0.81$\pm$0.34  &  56.1 $\pm$12.0    &   	1.54$\pm$0.30  &  34.3 $\pm$5.7  \\	       
61      &  3044856217706537088 &  0.33    &    	3.59$\pm$0.38  &  126.8$\pm$3.1	 &   	  -      &       -        &     	0.47$\pm$0.12  &  10.9 $\pm$7.3     &   	1.02$\pm$0.04  &  11.7 $\pm$1.2  \\       	
62      &  3044668579176599168 &  0.35    &    	1.06$\pm$0.10  &  15.6 $\pm$2.7	 &     0.58$\pm$0.02   &  179.8$\pm$1.1     &         -   &              - &   	0.53$\pm$0.07  &  12.9 $\pm$3.6  \\	       
63      &  3044668407377917184 &  0.34    &    	0.83$\pm$0.15  &  14.1 $\pm$5.0	 &     1.02$\pm$0.01   &  176.5$\pm$0.1     &       0.73$\pm$0.10  &  166.8$\pm$4.0     &   	1.13$\pm$0.01  &  173.1$\pm$0.4  \\	      
64*      &  3044667720183159168 &  0.12    &    	0.78$\pm$0.29  &  175.8$\pm$10.5	 &     1.20$\pm$0.08   &  170.7$\pm$1.8     &          -     &      -         &   	1.35$\pm$0.39  &  119.9$\pm$8.5  \\	      
65      &  3044856286426008320 &  0.36    &    	1.04$\pm$0.33  &  171.1$\pm$9.1	 &   	  -      &      -         &      	    -     &              - &   	1.00$\pm$0.22  &  10.3 $\pm$6.4  \\       		  
66*      &  3044857832614228608 &  0.66    &    	2.40$\pm$0.18  &  126.4$\pm$2.1	 &     1.44$\pm$0.62   &  139.6$\pm$12.3    &       1.58$\pm$0.09  &  132.4$\pm$1.7     &   	1.68$\pm$0.51  &  147.0$\pm$8.6  \\	     
67      &  3044858004412908288 &  1.83    &    	1.69$\pm$0.06  &  174.4$\pm$1.0	 &     1.76$\pm$0.07   &  174.0$\pm$1.2     &       1.76$\pm$0.12  &  166.0$\pm$1.9     &   	1.87$\pm$0.07  &  171.6$\pm$1.0  \\	     
68*      &  3044670125364791040 &  0.71    &    	2.35$\pm$0.27  &  154.8$\pm$3.3	 &     2.46$\pm$0.10   &  150.3$\pm$1.1     &       2.80$\pm$0.21  &  146.5$\pm$2.1     &   	2.60$\pm$0.06  &  153.1$\pm$0.7  \\	       	
69*      &  3044668304298684800 &  0.41    &    	1.39$\pm$0.01  &  1.0  $\pm$0.3	 &     1.92$\pm$0.29   &  166.1$\pm$4.4     &       1.66$\pm$0.05  &  165.2$\pm$0.9     &   	1.93$\pm$0.22  &  169.1$\pm$3.3  \\	      
70      &  3044667479664967808 &  0.47    &    	0.91$\pm$0.09  &  163.7$\pm$2.7	 &     1.03$\pm$0.06   &  168.6$\pm$1.8     &       0.56$\pm$0.12  &  135.7$\pm$6.1     &   	    - &  -        \\	       
71      &  3044669850486886912 &  0.69    &    	2.29$\pm$0.28  &  145.3$\pm$3.5	 &     1.56$\pm$0.01   &  149.2$\pm$0.1     &       2.05$\pm$0.18  &  141.9$\pm$2.5     &   	1.74$\pm$0.07  &  146.9$\pm$1.2  \\	       	
72*      &  3044667445305227136 &  0.56    &    	1.34$\pm$0.15  &  174.4$\pm$3.3	 &     1.59$\pm$0.35   &  166.9$\pm$6.2     &       1.62$\pm$0.06  &  168.8$\pm$1.1     &   	1.75$\pm$0.16  &  162.7$\pm$2.5  \\	     
73      &  3044858244931080320 &  2.59    &    	1.82$\pm$0.01  &  20.1 $\pm$0.1	 &     0.34$\pm$0.01   &  159.0$\pm$1.0     &       0.62$\pm$0.10  &  139.6$\pm$4.6     &   	1.07$\pm$0.41  &  8.5  $\pm$10.8 \\	       
74      &  3044857832614230528 &  0.62    &    	1.58$\pm$0.06  &  145.9$\pm$1.1	 &     2.73$\pm$0.06   &  133.7$\pm$0.6     &       2.62$\pm$0.25  &  136.4$\pm$2.8     &   	2.01$\pm$0.17  &  142.3$\pm$2.4  \\	       	
75      &  3044857866973962624 &  0.70    &    	    -     &    -   	 &     1.79$\pm$0.12   &  150.3$\pm$1.9     &       1.32$\pm$0.08  &  140.4$\pm$1.7     &   	2.48$\pm$0.78  &  139.2$\pm$9.1  \\        
76      &  3044668441737656192 &  0.12    &    	   -     &     -  	 &     2.55$\pm$0.51   &  31.7 $\pm$5.7     &          -     &     - &   	1.04$\pm$0.27  &  25.7 $\pm$7.4  \\        
77      &  3044669747407687808 &  0.87    &    	   -     &    -   	 &     3.18$\pm$0.04   &  153.1$\pm$0.3     &       3.59$\pm$0.16  &  140.8$\pm$1.3     &   	3.48$\pm$0.05  &  143.6$\pm$0.4  \\       
78      &  3044668132500011136 &  0.15    &    	    -     &   -    	 &     1.74$\pm$0.06   &  170.1$\pm$1.0     &       0.89$\pm$0.05  &  150.3$\pm$1.6     &   	1.16$\pm$0.11  &  159.2$\pm$2.7  \\       
79      &  3044856355145484928 &  0.13    &    	   -     &   -    	 &     5.37$\pm$0.88   &  127.0$\pm$4.7     &       1.20$\pm$0.11  &  145.2$\pm$2.7     &   	1.80$\pm$0.04  &  138.3$\pm$0.6  \\       
80      &  3044665349361182208 &  0.70    &    	0.77$\pm$0.12  &  3.7  $\pm$4.7 	 &  -	       &      -        &       0.43$\pm$0.05  &  122.9$\pm$3.4     &   	-        &  -       \\		
81*      &  3044662974233986176 &  0.16    &    	0.69$\pm$0.02  &  149.4$\pm$0.8	 &  	 -      &      -       &      	0.74$\pm$0.14  &  156.2$\pm$5.5     &   	-         &  -        \\      		 
82*      &  3044664146770344832 &  0.49    &    	0.74$\pm$0.07  &  162.2$\pm$2.6	 &     0.62$\pm$0.01   &  7.3  $\pm$0.6     &       0.64$\pm$0.17  &  152.6$\pm$7.8     &   	 -        &  -        \\	     
83      &  3044664112410613504 &  0.54    &    	1.20$\pm$0.32  &  169.1$\pm$7.6	 &     0.42$\pm$0.16   &  14.4 $\pm$10.9    &    	1.15$\pm$0.05  &  154.0$\pm$1.2     &   	1.43$\pm$0.13  &  158.1$\pm$2.6  \\
84*      &  3044664009331409152 &  0.46    &    	0.57$\pm$0.02  &  166.0$\pm$0.9	 &     1.06$\pm$0.04   &  174.1$\pm$1.0     &       0.90$\pm$0.32  &  13.1 $\pm$10.1    &   	    - &  -       \\	     
85      &  3044663356496399360 &  0.13    &    	0.50$\pm$0.17  &  40.2 $\pm$9.8	 &     0.41$\pm$0.08   &  21.0 $\pm$5.4     &       0.34$\pm$0.09  &  144.2$\pm$7.2     &   	   - &  -         \\	      
86      &  3044663283475304576 &  0.30    &    	1.02$\pm$0.07  &  4.6  $\pm$2.1	 &     0.92$\pm$0.02   &  169.7$\pm$0.8     &       0.94$\pm$0.01  &  159.3$\pm$0.4     &   	    - &  -         \\	    
87      &  3044663974971675008 &  0.37    &    	1.24$\pm$0.15  &  169.3$\pm$3.4	 &     1.88$\pm$0.14   &  164.0$\pm$2.2     &       1.21$\pm$0.17  &  161.5$\pm$4.1     &   	1.86$\pm$0.43  &  160.3$\pm$6.5  \\	      
88      &  3044667136067593856 &  1.24    &    	0.17$\pm$0.08  &  172.0$\pm$13.1 &     0.54$\pm$0.02   &  13.3 $\pm$1.0     &           -     &       -        &   	0.95$\pm$0.19  &  132.5$\pm$5.8  \\	     
89      &  3044663597014559232 &  0.35    &    	0.89$\pm$0.07  &  174.0$\pm$2.1	 &     1.13$\pm$0.34   &  172.1$\pm$8.5     &       0.91$\pm$0.14  &  165.7$\pm$4.4     &   	1.23$\pm$0.11  &  166.7$\pm$2.5  \\	    
90*      &  3044663631374306176 &  0.41    &    	1.47$\pm$0.26  &  169.4$\pm$5.2	 &     1.15$\pm$0.09   &  173.3$\pm$2.3     &       1.09$\pm$0.09  &  162.2$\pm$2.4     &   	1.22$\pm$0.10  &  169.3$\pm$2.4  \\	      
91*      &  3044667204787065216 &  0.18    &    	1.52$\pm$0.03  &  174.3$\pm$0.5	 &     0.69$\pm$0.18   &  170.4$\pm$7.3     &       0.82$\pm$0.18  &  112.4$\pm$6.2     &   	-         &  -       \\	     
92      &  3044667028683111296 &  0.40    &    	1.23$\pm$0.18  &  135.9$\pm$4.1	 &     0.99$\pm$0.03   &  179.8$\pm$0.8     &       1.13$\pm$0.09  &  171.0$\pm$2.2     &   	-        &  -      \\	    
93      &  3044666616366252160 &  0.43    &    	1.38$\pm$0.06  &  13.5 $\pm$1.1	 &     0.88$\pm$0.13   &  175.5$\pm$4.1     &       0.42$\pm$0.02  &  160.9$\pm$1.6     &   	    - &  -       \\	    
94      &  3044666689391008768 &  0.19    &    	    -     &     -           &     0.82$\pm$0.27   &  180.0$\pm$9.5     &        -     &             - &   	0.50$\pm$0.05  &  176.1$\pm$2.7  \\	      
95      &  3044663493935369600 &  0.20    &    	    -     &     -  	 &     0.18$\pm$0.07   &  124.3$\pm$12.1    &       0.46$\pm$0.14  &  7.2  $\pm$8.5     &   	1.22$\pm$0.07  &  24.3 $\pm$1.6  \\	    
96      &  3044667337920755456 &  0.23    &    	    -     &      - 	 &     4.75$\pm$1.18   &  24.3 $\pm$7.1     &       0.58$\pm$0.02  &  167.1$\pm$1.0     &   	0.37$\pm$0.10  &  17.7 $\pm$8.1  \\	    
97      &  3044572681146673536 &  0.34    &    	    -     &  -     	 &     0.69$\pm$0.04   &  164.9$\pm$1.5     &       -         &       -     &   	0.73$\pm$0.03  &  157.3$\pm$1.1  \\	     
98      &  3044667307866285568 &  0.41    &    	    -     &     -  	 &     2.03$\pm$0.01   &  167.0$\pm$0.1     &       1.88$\pm$0.02  &  174.4$\pm$0.2     &   	1.85$\pm$0.01  &  177.5$\pm$0.1  \\	      
99      &  3044663219057463808 &  0.53    &    	    -     &     -  	 &     0.68$\pm$0.06   &  168.9$\pm$2.5     &       1.33$\pm$0.14  &  166.9$\pm$2.9     &   	    -      &  -     \\	      
100*     &  3044669506889464320 &  0.16    &    	    -     &     -  	 &  	   -     &              - &       1.33$\pm$0.09  &  153.9$\pm$2.0     &   	1.38$\pm$0.14  &  163.2$\pm$2.9  \\	    
101*     &  3044669232011557760 &  0.16    &    	    -     &     -  	 &  	  -      &              - &    	1.78$\pm$0.10  &  154.8$\pm$1.6     &   	1.51$\pm$0.13  &  160.3$\pm$2.5  \\      	
102*     &  3044669300731031680 &  0.35    &    	    -     &    -   	 &  	  -      &              - &    	1.42$\pm$0.12  &  155.2$\pm$2.5     &   	1.24$\pm$0.28  &  166.6$\pm$6.5  \\      	
103*     &  3044665830397496576 &  0.16    &     1.21 $\pm$0.07  &  141.1$\pm$1.5	 &  	  -      &      -         &           -     &              - &         -     &     -       \\  
104*     &  3044668854054467840 &  1.42    &     1.43 $\pm$0.33  &  153.9$\pm$6.6	 &  	  -      &      -         &           -     &              - &         -     &       -     \\  
105*     &  3044669884846621824 &  0.12    &     2.05 $\pm$0.28  &  145.6$\pm$4.0	 &    	  -      &        -       &           -     &              - &         -     &     -       \\  
106*     &  3044669987925820416 &  0.22    &     1.64 $\pm$0.02  &  128.0$\pm$0.3	 &  	  -      &      -         &           -     &              - &         -     &     -       \\  
107*     &  3044670297163453824 &  0.37    &     1.34 $\pm$0.58  &  150.6$\pm$12.2	 &  	  -      &        -       &           -     &              - &         -     &      -      \\  
108*     &  3044669919206348416 &  0.35    &     1.95 $\pm$0.12  &  129.8$\pm$1.9	 &  	  -      &       -        &           -     &              - &    	    -     &      -      \\  
109     &  3044864532763156992 &  0.40    &             - &      -        &  	 -      &    -         &    	0.30$\pm$0.12  &  174.8$\pm$11.5    &   	   -     &      -      \\
110     &  3044675622913302016 &  3.77    &     2.52 $\pm$0.65  &  10.0 $\pm$7.3      &  	  -      &      -         &    	0.59$\pm$0.10  &  159.6$\pm$4.8     &     0.85$\pm$0.16  &  1.8  $\pm$5.5  \\
111     &  3044856423864969088 &  0.24    &     1.13 $\pm$0.23  &  160.3$\pm$5.9 	 &    	  -      &       -        &        -        &              - &         -     &       -    \\  
112*     &  3044668029420812800 &  0.44    &     0.96 $\pm$0.33  &  69.8 $\pm$9.8 	 &  	  -      &      -         &       -         &              - &         -     &     -   \\  
113*     &  3044855736670207744 &  0.44    &     0.81 $\pm$0.08  &  178.9$\pm$2.9 	 &    	  -      &      -         &       -         &              - &         -     &       -    \\  
114     &  3044856217706537472 &  0.31    &     2.74 $\pm$0.65  &  176.9$\pm$6.8 	 &    	  -      &     -          &     -           &              - &   	    -     &       -     \\  
115     &  3044670052342416512 &  0.46    &  	        - &      -      	 &      0.9$\pm$0.02   &  6.0  $\pm$0.7     &    	    -     &       -        &     0.91$\pm$0.26  &  10.0 $\pm$8.2  \\
116*     &  3044670022285588224 &  0.51    &  	2.90$\pm$0.37  &  138.1$\pm$3.7 	 &  	  -       &  -            &           -     &      -        &   	    -     &       -     \\  
117*     &  3044665589879322112 &  0.21    &     0.82 $\pm$0.01  &  122.7$\pm$0.5 	 &  -	         &  -            &           -     &      - &         -     &       -     \\  
118*     &  3044665796037749632 &  0.33    &     1.31 $\pm$0.32  &  148.5$\pm$6.9 	 &  -	         &  -            &           -     &      - &      -        &      -     \\  
119     &  3044665761678261760 &  0.30    &  	1.33$\pm$0.10  &  145.4$\pm$2.1 	 &	-    &  -                      &   -           &   	    -     &      -  &-   \\  
120     &  3044663115978197504 &  0.15    &     0.69 $\pm$0.06  &  18.0 $\pm$2.3 	 &  -	         &  -            &           -     &       - &         -     &      -     \\  
121     &  3044662703661353600 &  0.55    &     1.32 $\pm$0.23  &  145.7$\pm$4.9 	 &   - 	         &  -            &           -     &       - &         -     &       -     \\  
122     &  3044662596280537600 &  0.88    &     1.45 $\pm$0.08  &  174.2$\pm$1.5 	 &  -  	         &  -            &           -     &      - &         -     &       -    \\  
123*     &  3044662909819779840 &  0.11    &     1.14 $\pm$0.21  &  164.6$\pm$5.3 	 &  -  	         &  -            &           -     &       -        &         -     &       -    \\  
124*     &  3044662974235708928 &  0.05    &  	       - &      -    	 &-	   &  -        &    	0.37$\pm$0.07  &  145.0$\pm$5.8     &   -	        &     -      \\
125     &  3044662875460051200 &  1.14    &     1.79 $\pm$0.38  &  106.9$\pm$6.1	 &   - 	         &  -            &           -     &      - &         -     &       -     \\  
126*     &  3044662871159883648 &  0.53    &  	2.83$\pm$1.32  &  93.1 $\pm$13.3	 &	 - &  -           & -           &      - &   	    -     &      -     \\  
127*     &  3044663665734034688 &  0.36    &  	0.79$\pm$0.24  &  97.2 $\pm$8.6	 &  -	         &  -            &           -     &       -        &   	   -     &      -     \\  		
128*     &  3044668643601838464 &  0.50    &  	1.25$\pm$0.19  &  150.1$\pm$4.3	 &  	  -      &  -            &           -     &       - &   	    -     &      -     \\  
129     &  3044663768813248640 &  0.26    &          -      &       -	 &    	 -      &  -          &    	0.51$\pm$0.12  &  105.2$\pm$6.6     &         -     &      -     \\
130     &  3044663768813249536 &  0.29    &     1.64 $\pm$0.21  &  169.3$\pm$3.7      &  	  -       &  -            &          -     &       - &         -     &       -     \\  
131     &  3044667582744222080 &  0.33    &          -     &       -        &  	0.36$\pm$0.07   &  14.2 $\pm$5.3     &    	   -     &       - &         -     &       -    \\
132     &  3044666826820729088 &  0.15    &  	    -     &       -         &  	0.83$\pm$0.02   &  3.5  $\pm$0.8     &    	    -     &       -      &   	    -     &       -     \\
133     &  3044666856886407168 &  0.17    &          -     &       -         &  	0.64$\pm$0.16   &  169.4$\pm$7.2     &    	    -     &       - &         -     &      -    \\
134     &  3044666723750771968 &  0.10    &  	    -     &       -         &       1.06$\pm$0.39   &  167.3$\pm$10.4    &    	    -     &       -        &   	    -     &       -    \\
135     &  3044663386552561536 &  0.33    &  	    -      &       -         &         - &  -          &    	0.42$\pm$0.19  &  158.0$\pm$13.0    &   	    -     &       -     \\		                                    \hline	
&&&&& Field region \\
\hline
136     &  3044619994505703680 &  0.24    &    	0.99$\pm$0.24  &  157.6$\pm$7.0	 &     1.47$\pm$0.68   &  153.8$\pm$13.1    &      	1.29$\pm$0.54  &  142.7$\pm$12.0    &   	1.45$\pm$0.07  &  151.3$\pm$1.4  \\      	
137     &  3044618444014605568 &  0.30    &    	1.25$\pm$0.14  &  129.9$\pm$3.3	 &     1.93$\pm$0.04   &  156.1$\pm$0.5     &      	2.20$\pm$0.22  &  133.3$\pm$2.9     &   	1.57$\pm$0.06  &  139.4$\pm$1.1  \\     
138     &  3044619685268066432 &  0.11    &    	0.78$\pm$0.17  &  145.0$\pm$6.1	 &     1.54$\pm$0.34   &  152.2$\pm$6.2     &      	1.05$\pm$0.15  &  142.1$\pm$4.0     &   	0.61$\pm$0.05  &  140.5$\pm$2.5  \\      		
139     &  3044619685268070656 &  0.17    &    	1.14$\pm$0.52  &  150.5$\pm$12.9	 &     1.06$\pm$0.02   &  153.3$\pm$0.5     &      	1.62$\pm$0.10  &  138.9$\pm$1.8     &   	1.01$\pm$0.08  &  139.4$\pm$2.4  \\       		 
140     &  3044619616548784512 &  0.05    &    	1.36$\pm$0.18  &  176.5$\pm$3.9	 &     1.04$\pm$0.21   &  163.4$\pm$5.7     &      	1.65$\pm$0.31  &  143.7$\pm$5.3     &   	0.87$\pm$0.10  &  138.1$\pm$3.4  \\      
141     &  3044619376030443136 &  0.14    &    	2.95$\pm$0.27  &  147.8$\pm$2.6	 &     1.74$\pm$0.38   &  147.5$\pm$6.2     &      	1.87$\pm$0.16  &  135.5$\pm$2.5     &   	1.92$\pm$0.44  &  140.5$\pm$6.6  \\       
142     &  3044623018162683136 &  0.20    &    	2.25$\pm$0.29  &  145.7$\pm$3.7	 &     2.37$\pm$0.05   &  146.1$\pm$0.6     &      	2.00$\pm$0.21  &  139.2$\pm$3.0     &   	1.66$\pm$0.34  &  138.8$\pm$5.8  \\       		   
143     &  3044623086882160256 &  0.27    &    	2.20$\pm$0.42  &  149.6$\pm$5.4	 &     1.27$\pm$0.01   &  163.6$\pm$0.1     &      	0.91$\pm$0.21  &  140.6$\pm$6.5     &   	0.95$\pm$0.22  &  145.6$\pm$6.6  \\       		
144     &  3044619582188871808 &  0.09    &    	1.58$\pm$0.01  &  146.4$\pm$0.1	 &     1.74$\pm$0.13   &  144.3$\pm$2.1     &      	1.51$\pm$0.12  &  135.5$\pm$2.3     &   	1.33$\pm$0.14  &  146.9$\pm$3.0  \\       		
145     &  3044619479109661184 &  0.16    &    	    -      &  -    	 &     2.64$\pm$0.18   &  137.0$\pm$2.0     &    	2.44$\pm$0.16  &  133.2$\pm$1.9     &   	2.24$\pm$0.07  &  140.7$\pm$0.9  \\       		  
146     &  3044613363076252416 &  0.23    &    	0.72$\pm$0.21  &  102.6$\pm$8.1      &     1.32$\pm$0.27   &  145.1$\pm$5.8     &    	1.48$\pm$0.20  &  139.0$\pm$3.8     &   	1.16$\pm$0.36  &  150.3$\pm$9.0  \\       		   
147     &  3044623293040584064 &  0.36    &    	2.21$\pm$0.40  &  112.6$\pm$5.2      &     1.60$\pm$0.22   &  146.8$\pm$4.0     &    	1.90$\pm$0.14  &  139.9$\pm$2.1     &   	1.97$\pm$0.30  &  149.4$\pm$4.4  \\       		   
148     &  3044622915083470464 &  0.29    &    	0.15$\pm$0.03  &  161.5$\pm$6.0      &     0.40$\pm$0.08   &  54.0 $\pm$5.7     &    	    -     &       -       &   	    -     &      -     \\       		  
149     &  3044622708925050496 &  0.06    &    	1.59$\pm$0.04  &  136.3$\pm$0.6      &     1.06$\pm$0.15   &  129.1$\pm$4.1     &    	0.70$\pm$0.05  &  137.5$\pm$1.9     &   	0.84$\pm$0.25  &  144.0$\pm$8.5  \\       
150     &  3044622880723747584 &  0.06    &    	3.83$\pm$0.05  &  142.6$\pm$0.4      &     1.99$\pm$0.53   &  128.5$\pm$7.7     &    	2.54$\pm$0.25  &  129.3$\pm$2.8     &   	2.09$\pm$0.13  &  136.9$\pm$1.8  \\       
151     &  3044619513469410176 &  0.39    &    	    -     &       -	 &     3.80$\pm$0.23   &  143.7$\pm$1.7     &       2.71$\pm$0.10  &  132.5$\pm$1.1     &   	1.77$\pm$0.28  &  127.8$\pm$4.6  \\	      
152     &  3044616592891655808 &  0.20    &    	2.13$\pm$0.28  &  18.5 $\pm$3.7      &     0.54$\pm$0.10   &  142.3$\pm$5.2     &    	1.02$\pm$0.39  &  124.6$\pm$10.9    &   	    -     &       -     \\       		
153     &  3044622949443216896 &  0.29    &    	0.73$\pm$0.11  &  112.3$\pm$4.3      &  	-       &       -        &    	0.89$\pm$0.09  &  116.8$\pm$2.8     &   	0.57$\pm$0.05  &  120.1$\pm$2.5  \\
154     &  3044623705357465216 &  0.15    &    	0.63$\pm$0.15  &  103.5$\pm$6.6      &     0.50$\pm$0.07   &  117.1$\pm$4.0     &    	0.55$\pm$0.22  &  114.5$\pm$11.4    &   	0.66$\pm$0.13  &  159.4$\pm$5.7  \\       	
155     &  3044623808436866560 &  0.76    &    	1.85$\pm$0.16  &  133.7$\pm$2.5      &     1.57$\pm$0.05   &  127.3$\pm$1.0     &    	1.33$\pm$0.16  &  137.8$\pm$3.5     &   	0.78$\pm$0.36  &  143.3$\pm$13.2 \\      		
156     &  3044623602278261120 &  0.09    &    	   -     &      -   	 & 	-       &       -       &     	0.67$\pm$0.15  &  129.7$\pm$6.3     &   	1.17$\pm$0.06  &  154.0$\pm$1.5  \\		  
157     &  3044622812004295936 &  0.34    &    	1.76$\pm$0.49  &  127.0$\pm$8.1      &     0.89$\pm$0.21   &  141.1$\pm$6.9     &    	1.01$\pm$0.20  &  146.1$\pm$5.6     &   	1.09$\pm$0.22  &  145.8$\pm$5.9  \\      	
158     &  3044616833409833984 &  0.19    &    	0.84$\pm$0.13  &  165.6$\pm$4.6      &     1.11$\pm$0.20   &  144.4$\pm$5.2     &    	0.88$\pm$0.16  &  147.2$\pm$5.3     &   	1.83$\pm$0.27  &  124.0$\pm$4.2  \\      		
159     &  3044616558531934720 &  1.91    &    	1.65$\pm$0.04  &  165.0$\pm$0.7      &     1.03$\pm$0.11   &  146.7$\pm$3.1     &    	1.60$\pm$0.02  &  130.4$\pm$0.3     &   	1.84$\pm$0.25  &  127.5$\pm$3.8  \\      	
160     &  3044613603594436224 &  0.24    &    	    -     &       -	 &     0.28$\pm$0.09   &  126.5$\pm$9.7     &           -     &       -      &   	1.42$\pm$0.65  &  174.1$\pm$13.1 \\	      
161     &  3044613603594428416 &  0.23    &      1.95$\pm$0.69  &  131.2$\pm$10.2	 &     0.89$\pm$0.12   &  149.3$\pm$3.9     &      	0.73$\pm$0.02  &  146.9$\pm$0.7     &   	1.83$\pm$0.32  &  134.9$\pm$5.0  \\    
162     &  3044622983802946176 &  0.58    &      1.69$\pm$0.36  &  151.0$\pm$6.1	 &     1.85$\pm$0.55   &  148.2$\pm$8.4     &      	1.94$\pm$0.16  &  138.2$\pm$2.3     &   	1.51$\pm$0.39  &  120.7$\pm$7.2  \\      
163     &  3044622468406914560 &  1.22    &    	3.69$\pm$0.96  &  174.6$\pm$7.4	 &     1.53$\pm$0.55   &  158.5$\pm$10.2    &      	2.35$\pm$0.30  &  133.7$\pm$3.7     &   	2.06$\pm$0.11  &  145.1$\pm$1.5  \\      
164     &  3044616661611133696 &  0.41    &    	0.95$\pm$0.42  &  139.0$\pm$12.6	 &     1.09$\pm$0.05   &  137.1$\pm$1.4     &      	1.08$\pm$0.17  &  129.5$\pm$4.4     &   	0.87$\pm$0.10  &  140.6$\pm$3.1  \\       	
165     &  3044616867769566464 &  0.29    &    	1.04$\pm$0.37  &  122.7$\pm$10.2	 &     0.54$\pm$0.01   &  132.2$\pm$0.6     &      	0.92$\pm$0.19  &  122.8$\pm$5.9     &   	0.40$\pm$0.17  &  139.9$\pm$12.1 \\       		   
166     &  3044626591575395840 &  1.24    &    	0.23$\pm$0.10  &  89.1 $\pm$11.2	 &  	  -      &       -        &       0.59$\pm$0.22  &  13.5 $\pm$10.5    &   	0.34$\pm$0.13  &  60.8 $\pm$10.8 \\      
167     &  3044626007459848576 &  1.25    &    	1.08$\pm$0.53  &  174.5$\pm$20.1	 &     1.47$\pm$0.56   &  136.8$\pm$11.3    &       2.09$\pm$0.19  &  172.9$\pm$2.2     &   	2.45$\pm$0.42  &  146.4$\pm$4.9  \\      
168     &  3044626591575396864 &  1.46    &    	2.23$\pm$0.41  &  43.4 $\pm$5.7	 &     1.84$\pm$0.56   &  68.9 $\pm$8.3     &       1.43$\pm$0.16  &  135.9$\pm$3.5     &   	0.88$\pm$0.38  &  26.6 $\pm$12.5 \\       
169     &  3044625973100121984 &  1.03    &    	1.34$\pm$0.28  &  142.2$\pm$6.1	 &     1.91$\pm$0.03   &  147.2$\pm$0.4     &       1.56$\pm$0.16  &  148.9$\pm$2.9     &   	1.20$\pm$0.35  &  141.7$\pm$8.4  \\        
170     &  3044626282338024576 &  1.05    &    	0.84$\pm$0.23  &  136.0$\pm$8.4	 &     1.75$\pm$0.38   &  159.9$\pm$5.9     &       1.59$\pm$0.01  &  153.5$\pm$0.2     &   	1.16$\pm$0.14  &  150.7$\pm$3.4  \\      
171     &  3044626247978028672 &  1.14    &    	0.65$\pm$0.07  &  116.7$\pm$3.4	 &  	  -      &       -        &       1.13$\pm$0.15  &  127.5$\pm$4.2     &   	    -     &       - \\        
172     &  3044626144898817024 &  1.02    &    	    -     &       -	 &     0.81$\pm$0.26   &  102.6$\pm$9.0     &      	1.48$\pm$0.68  &  119.0$\pm$14.7    &   	1.68$\pm$0.14  &  152.8$\pm$2.4  \\     
173     &  3044638480044865536 &  0.67    &    	    -     &       -	 &     3.38$\pm$0.12   &  136.0$\pm$1.1     &      	   -     &       -     &   	2.17$\pm$0.24  &  25.5 $\pm$3.1  \\       		
174     &  3044638720562671488 &  0.41    &    	    -     &      -	 &     2.28$\pm$0.29   &  141.3$\pm$3.7     &      	2.86$\pm$0.21  &  133.6$\pm$2.4     &   	1.97$\pm$0.09  &  157.4$\pm$1.3  \\       	
175     &  3044626037515508096 &  0.65    &   	2.07$\pm$0.30  &  178.8$\pm$3.9	 &     2.13$\pm$0.28   &  129.9$\pm$4.1     &       2.57$\pm$0.13  &  136.0$\pm$1.6     &   	2.01$\pm$0.09  &  133.7$\pm$1.3  \\	     
176     &  3044624560046528000 &  0.34    &   	0.50$\pm$0.03  &  147.3$\pm$1.6	 &     0.67$\pm$0.15   &  24.1 $\pm$6.7     &       0.34$\pm$0.02  &  138.8$\pm$1.4     &    	    -     &       -    \\	       
177     &  3044637891625009536 &  1.09    &   	    -     &       -	 &     0.48$\pm$0.14   &  157.3$\pm$8.1     &    	0.94$\pm$0.12  &  145.3$\pm$3.9     &     1.16$\pm$0.02  &  138.5$\pm$0.4  \\
178     &  3044638102087753856 &  0.88    &   	1.48$\pm$0.62  &  138.7$\pm$12.7	 &     1.86$\pm$0.06   &  148.4$\pm$0.9     &       1.00$\pm$0.26  &  121.3$\pm$8.3     &   	    -     &       -     \\	      
179     &  3044638754922419584 &  1.15    &   	1.77$\pm$0.16  &  108.9$\pm$2.7	 &     1.12$\pm$0.51   &  170.5$\pm$21.6    &       3.27$\pm$0.02  &  128.4$\pm$0.2     &   	1.22$\pm$0.53  &  163.0$\pm$12.6 \\	       
180     &  3044638067727676800 &  1.31    &   	0.90$\pm$0.29  &  154.8$\pm$8.6	 &     1.57$\pm$0.02   &  160.1$\pm$0.3     &       1.57$\pm$0.28  &  163.1$\pm$4.7     &   	1.25$\pm$0.35  &  148.1$\pm$8.1  \\	    
181     &  3044637895929343104 &  1.36    &   	6.74$\pm$0.10  &  142.9$\pm$0.4	 &     1.78$\pm$0.61   &  134.6$\pm$10.4    &      	1.18$\pm$0.06  &  117.2$\pm$1.7     &   	1.51$\pm$0.14  &  129.6$\pm$2.7  \\       		
182     &  3044624633070348288 &  0.77    &   	8.24$\pm$2.55  &  128.7$\pm$10.7	 &     1.15$\pm$0.07   &  133.3$\pm$1.9     &      	2.06$\pm$0.03  &  138.0$\pm$0.4     &   	0.94$\pm$0.20  &  133.1$\pm$6.0  \\       		
183     &  3044624529991370112 &  0.62    &   	    -     &      -	 &  	-        &       -        &       1.51$\pm$0.04  &  161.8$\pm$0.6     &   	1.50$\pm$0.15  &  152.4$\pm$2.9  \\ 	          
184     &  3044638273886101888 &  1.42    &   	0.75$\pm$0.02  &  130.8$\pm$0.8    	 &     1.35$\pm$0.19   &  144.6$\pm$4.2     &    	1.43$\pm$0.42  &  138.7$\pm$9.2     &   	0.86$\pm$0.03  &  158.6$\pm$1.0  \\
185     &  3044638342605574784 &  1.84    &   	3.33$\pm$1.23  &  72.5 $\pm$9.5    	 &     1.08$\pm$0.15   &  175.7$\pm$3.6     &    	2.51$\pm$0.74  &  111.6$\pm$9.1     &   	1.34$\pm$0.21  &  173.5$\pm$4.6  \\
186     &  3044638342605581056 &  1.97    &   	1.28$\pm$0.26  &  144.9$\pm$5.7    	 &     1.43$\pm$0.01   &  146.6$\pm$0.1     &    	1.38$\pm$0.06  &  154.8$\pm$1.1     &   	1.14$\pm$0.14  &  149.8$\pm$3.5  \\	
187     &  3044638991136633728 &  2.00    &   	0.41$\pm$0.13  &  47.7 $\pm$9.6    	 &     0.33$\pm$0.05   &  180.0$\pm$3.9     &    	0.26$\pm$0.10  &  23.1 $\pm$11.4    &   	0.41$\pm$0.16  &  179.7$\pm$11.6 \\		
188     &  3044638200871683968 &  1.49    &   	1.48$\pm$0.13  &  92.2 $\pm$2.5    	 &     0.41$\pm$0.08   &  31.9 $\pm$6.1     &    	0.23$\pm$0.01  &  41.2 $\pm$0.9     &   	0.77$\pm$0.05  &  47.3 $\pm$2.0  \\
189     &  3044636658978434688 &  0.78    &   	1.13$\pm$0.38  &  147.8$\pm$9.4    	 &     1.62$\pm$0.35   &  141.8$\pm$6.3     &    	2.56$\pm$0.58  &  149.1$\pm$6.5     &   	1.84$\pm$0.25  &  139.9$\pm$3.9  \\	
190     &  3044636418460269184 &  0.67    &   	0.99$\pm$0.19  &  49.1 $\pm$5.8    	 &     1.03$\pm$0.32   &  81.6 $\pm$8.5     &    	1.01$\pm$0.13  &  105.0$\pm$3.6     &   	0.42$\pm$0.10  &  157.9$\pm$6.9  \\
191     &  3044636727697909376 &  0.82    &   	2.62$\pm$0.82  &  62.5 $\pm$8.3    	 &     2.32$\pm$0.10   &  70.7 $\pm$1.1     &    	1.09$\pm$0.05  &  29.3 $\pm$1.6     &   	1.36$\pm$0.14  &  64.0 $\pm$3.0  \\		
192     &  3044636452819996032 &  1.29    &   	1.06$\pm$0.03  &  62.4 $\pm$0.7    	 &     0.63$\pm$0.14   &  29.2 $\pm$6.6     &    	1.06$\pm$0.11  &  11.3 $\pm$2.8     &   	0.42$\pm$0.20  &  41.0 $\pm$13.4 \\	
193     &  3044636315381068544 &  1.04    &   	1.73$\pm$0.10  &  168.6$\pm$1.6    	 &     1.92$\pm$0.94   &  33.9 $\pm$14.6    &    	1.70$\pm$0.10  &  13.7 $\pm$1.6     &   	0.80$\pm$0.05  &  171.3$\pm$1.7  \\	
194     &  3044626110539088896 &  1.07    &   	    -     &       -       	 &  	-        &      -        &    	3.43$\pm$0.88  &  19.7 $\pm$7.7     &   	1.77$\pm$0.05  &  120.5$\pm$0.8  \\		
195     &  3044637861569264000 &  0.84    &   	3.53$\pm$1.21  &  157.9$\pm$8.7	 &  	-       &       -       &    	1.98$\pm$0.17  &  176.9$\pm$2.1     &   	1.28$\pm$0.57  &  153.5$\pm$19.5 \\
196     &  3044636349740792320 &  0.73    &  	1.35$\pm$0.24  &  137.8$\pm$5.4    	 &     1.64$\pm$0.20   &  131.4$\pm$3.7     &    	1.53$\pm$0.11  &  135.3$\pm$2.3     &   	0.97$\pm$0.13  &  139.6$\pm$3.8  \\
197     &  3044624598710622464 &  0.57    &    	- &	-			 &     1.38$\pm$0.66   &  10.4 $\pm$13.7	&	-		& - &   	0.99$\pm$0.21  &  132.9$\pm$6.0  \\
   
\hline
\end{longtable}
\footnotesize{ID marked with asterisk symbol (*) are members of the cluster NGC 2345.}
}

\begin{figure*}
\centering 
\begin{subfigure} {0.49\columnwidth} 
\centering 
\includegraphics[trim=1.8cm 0.5cm 1.8cm 0.1cm, width=\columnwidth]{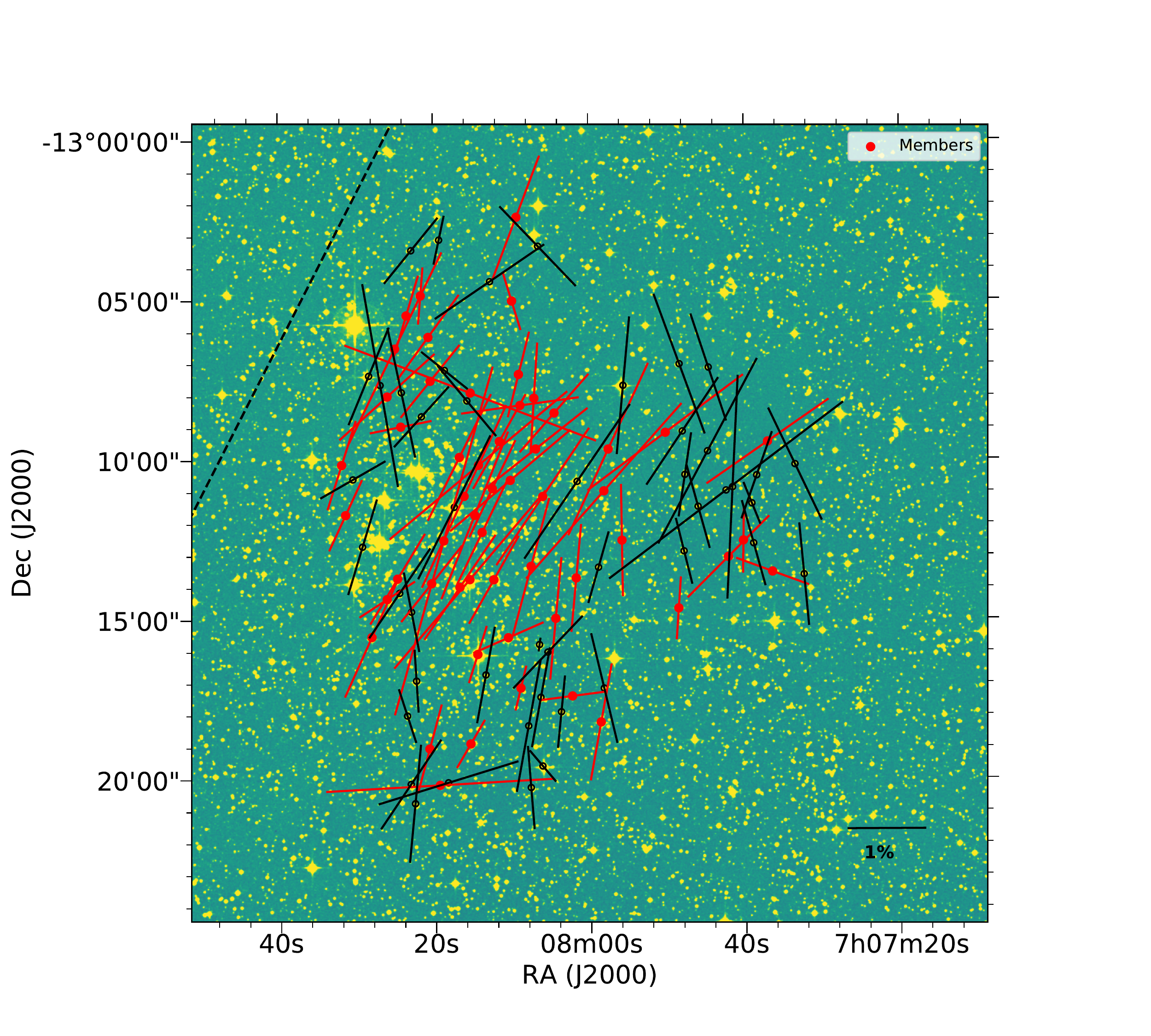}
\caption{B-Band} \label{fig:dss_M_B}  
\end{subfigure}
\begin{subfigure} {0.49\columnwidth} 
\centering 
\includegraphics[trim=1.8cm 0.5cm 1.8cm 0.1cm, width=\columnwidth]{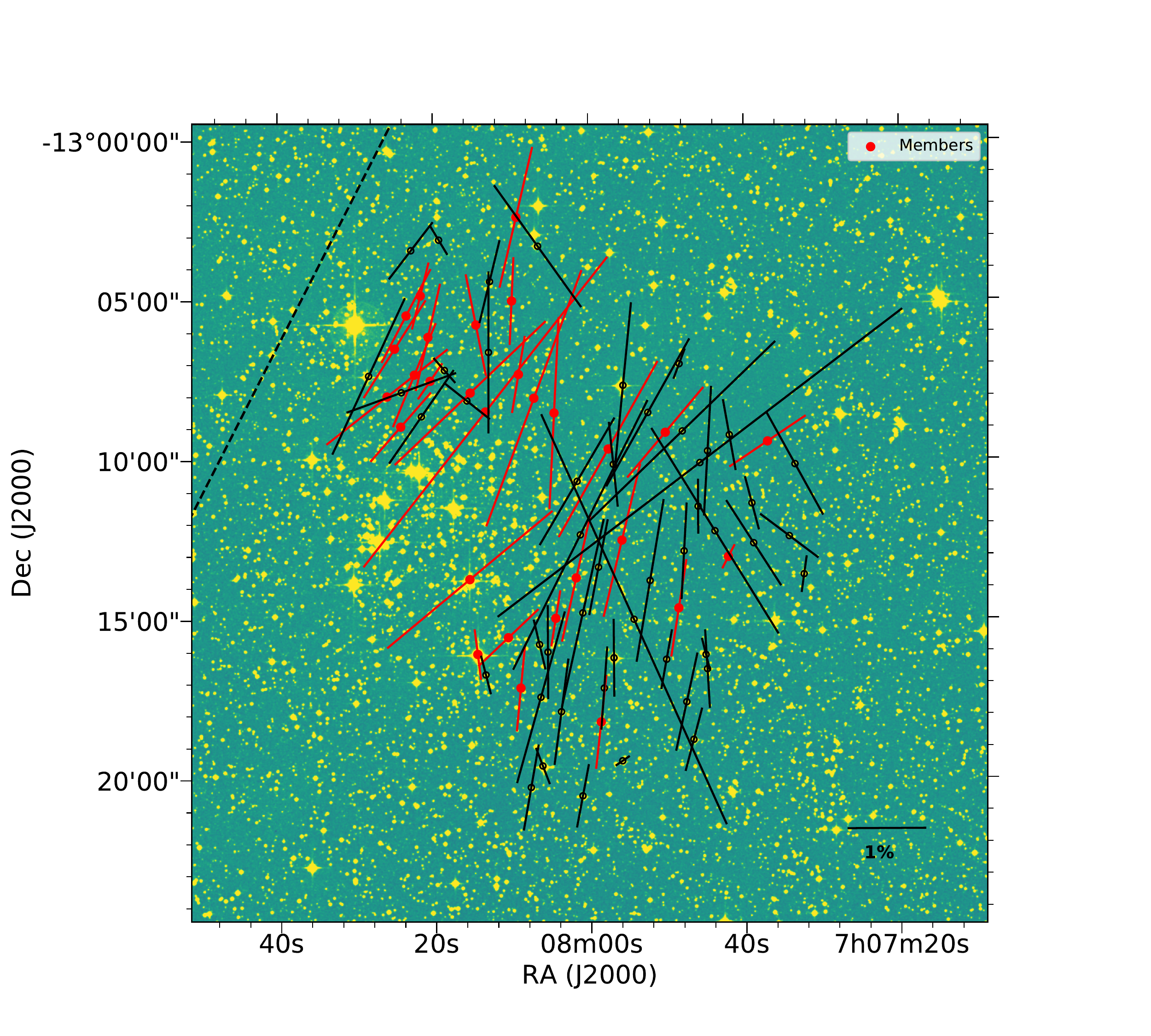}
\caption{V-Band} \label{fig:dss_M}  
\end{subfigure}
\begin{subfigure} {0.49\columnwidth} 
\centering 
\includegraphics[trim=1.8cm 0.5cm 1.8cm 0.1cm, width=\columnwidth]{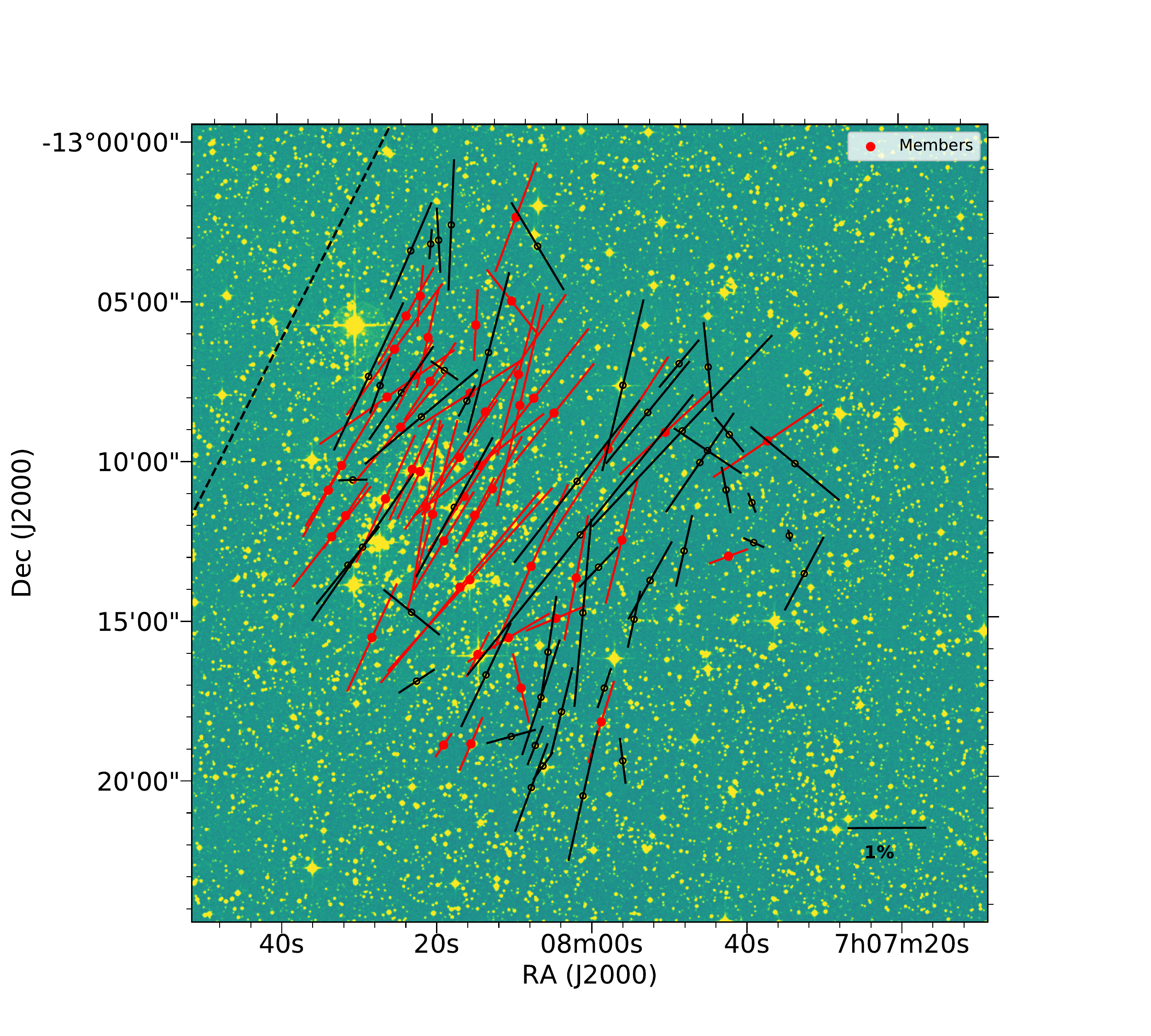}
\caption{R-Band} \label{fig:dss_M_R}  
\end{subfigure}
\begin{subfigure} {0.49\columnwidth} 
\centering 
\includegraphics[trim=1.8cm 0.5cm 1.8cm 0.1cm, width=\columnwidth]{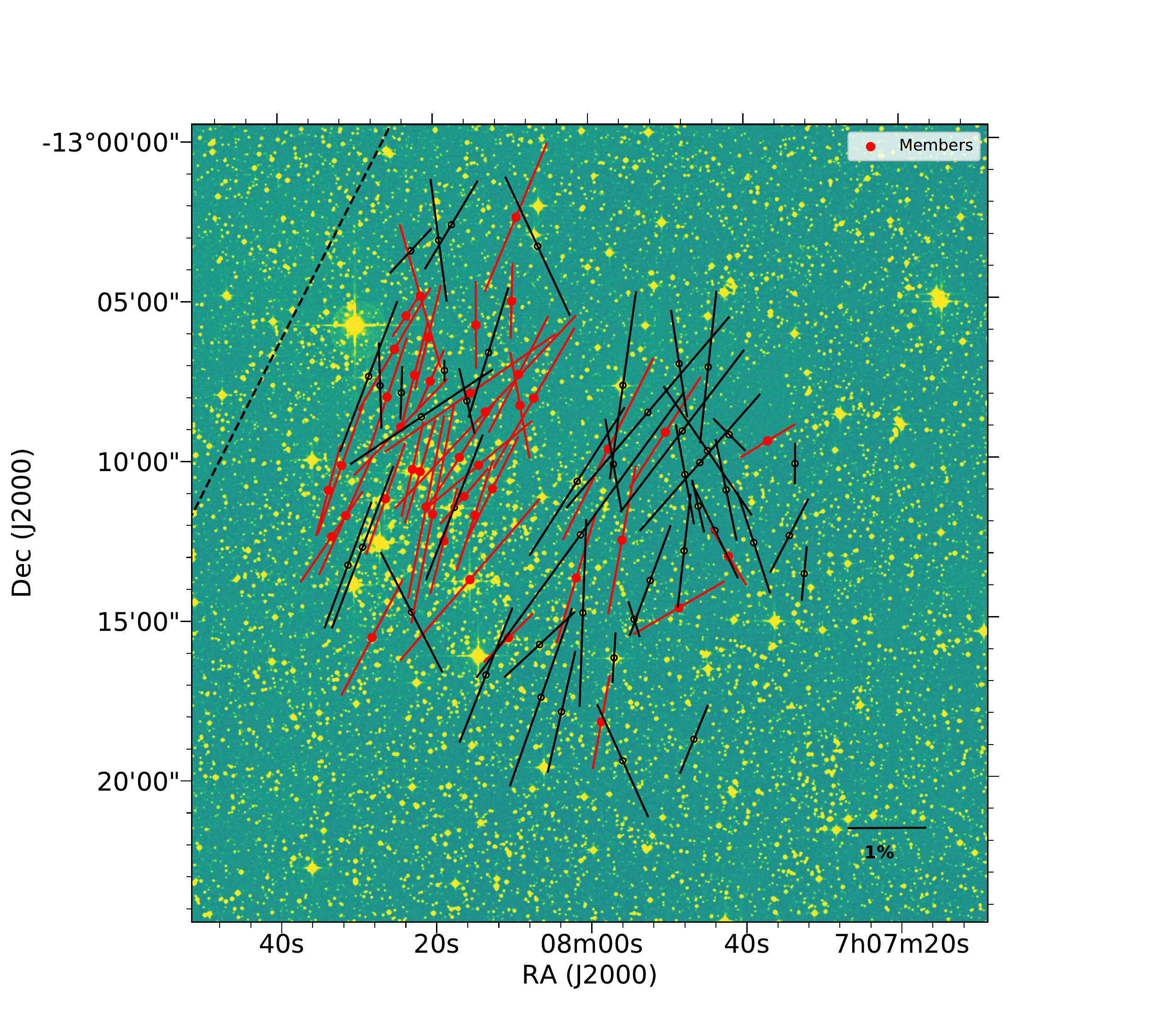}
\caption{I-Band} \label{fig:dss_M_I}  
\end{subfigure}
\caption{The polarization vectors in $B$, $V$, $R$, and $I$ bands are overplotted on 25\arcmin$\times$25\arcmin\ DSS $R$-band image of the cluster NGC 2345. The length of the vector shows the degree of polarization and their tilt denotes the position angle. The reference length for 1\% polarization is shown at the bottom right. The dotted line denotes the orientation of the projection of Galactic Parallel at the region. }
\label{fig:dss}
\end{figure*}

\begin{figure*}
\begin{subfigure} {0.49\columnwidth} 
\centering 
\includegraphics[trim=1.8cm 0.5cm 1.8cm 0.1cm, width=\columnwidth]{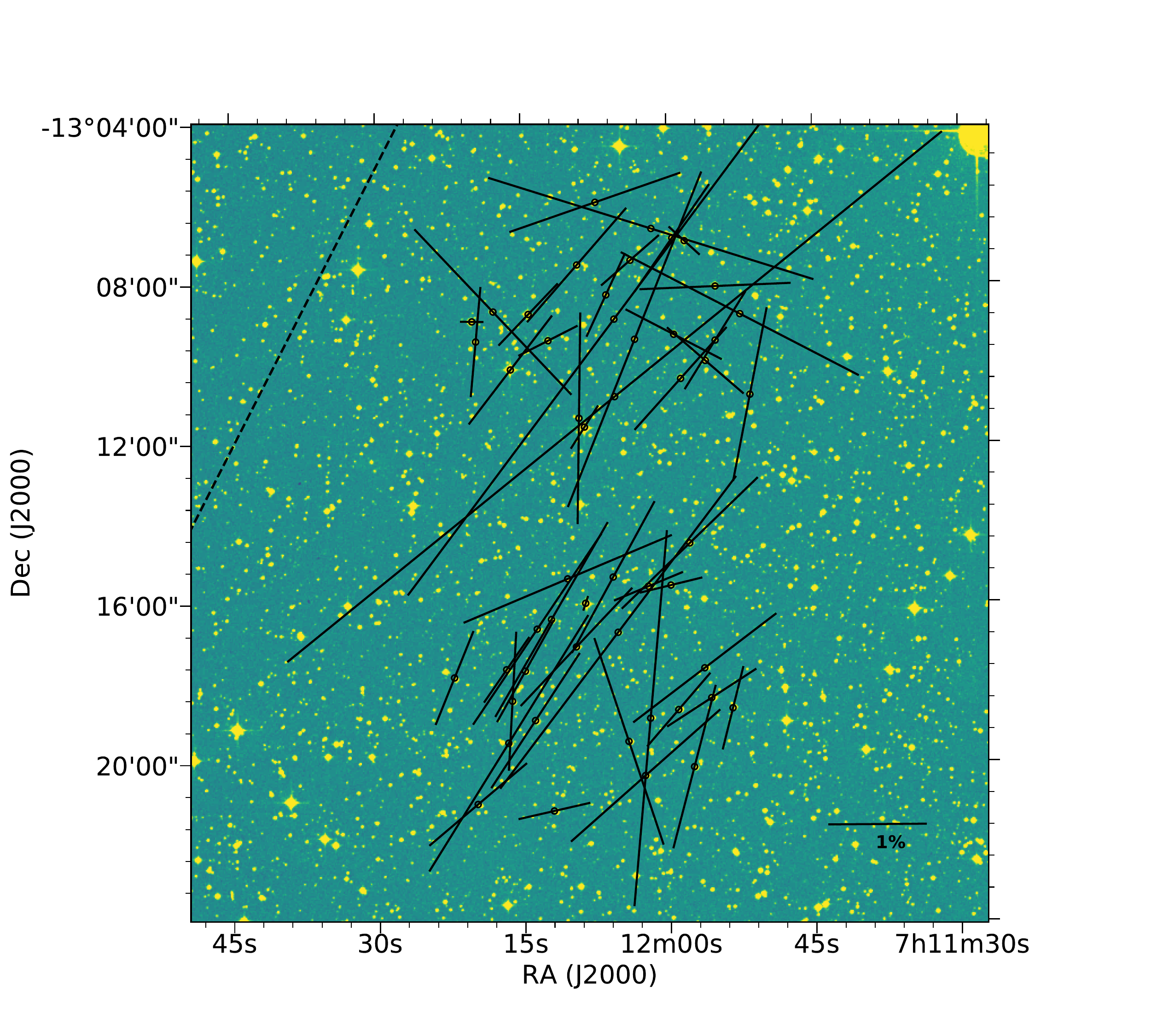}
\caption{B-Band} \label{fig:dss_F_B}  
\end{subfigure}
\begin{subfigure} {0.49\columnwidth} 
\centering 
\includegraphics[trim=1.8cm 0.5cm 1.8cm 0.1cm, width=\columnwidth]{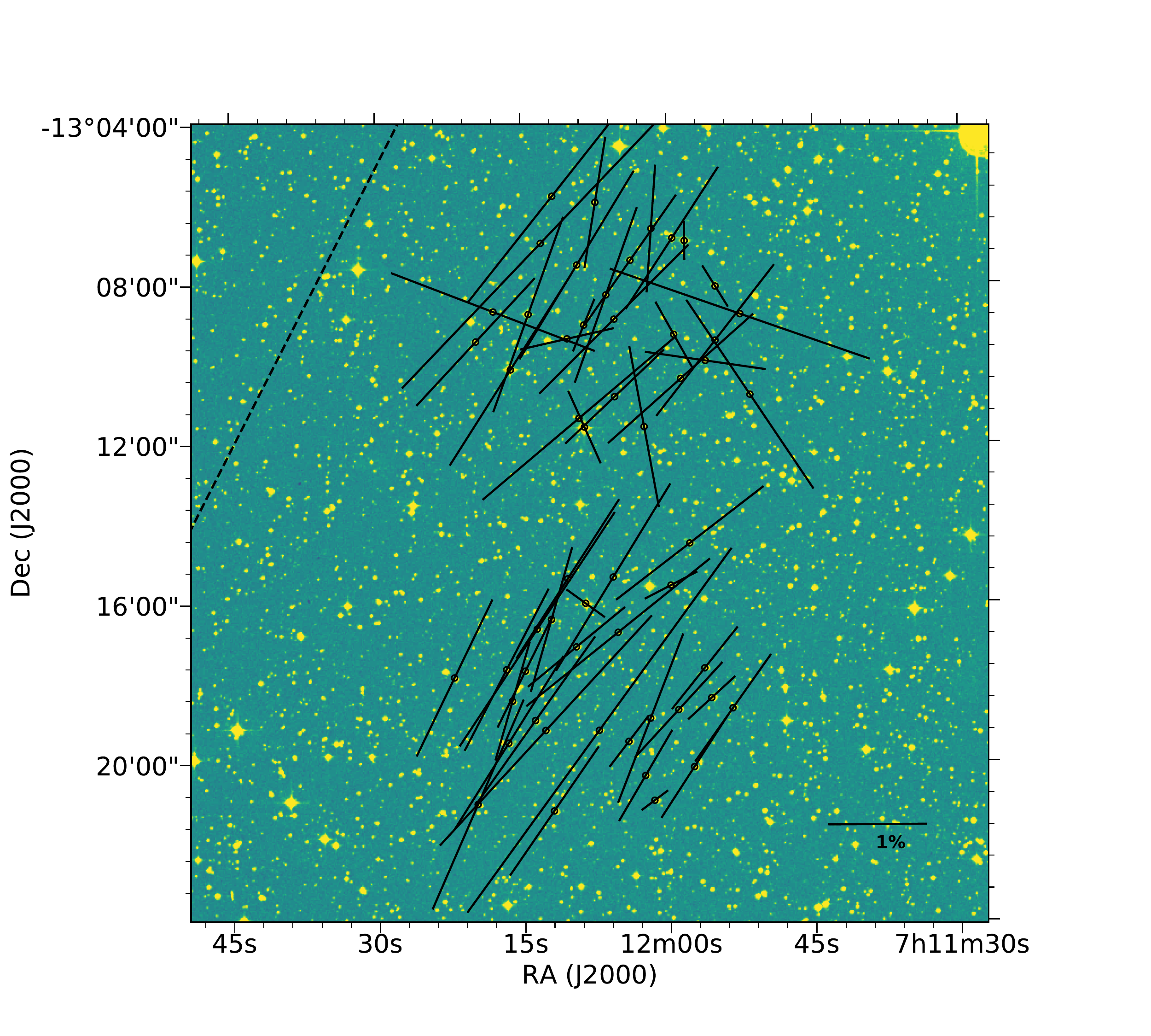}
\caption{V-Band} \label{fig:dss_F}  
\end{subfigure}
\begin{subfigure} {0.49\columnwidth} 
\centering 
\includegraphics[trim=1.8cm 0.5cm 1.8cm 0.1cm, width=\columnwidth]{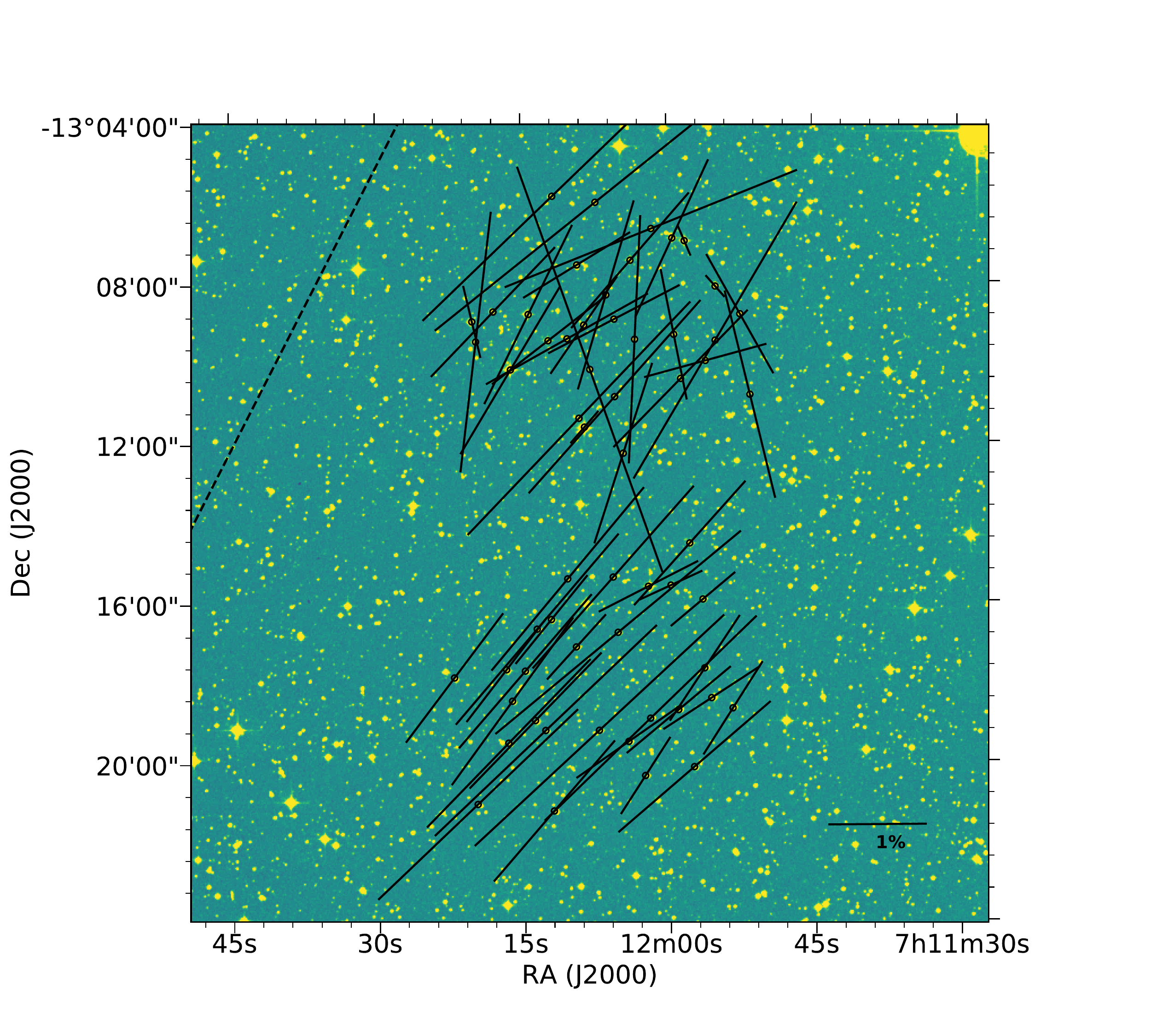}
\caption{R-Band} \label{fig:dss_F_R}  
\end{subfigure}
\begin{subfigure} {0.49\columnwidth} 
\centering 
\includegraphics[trim=1.8cm 0.5cm 1.8cm 0.1cm, width=\columnwidth]{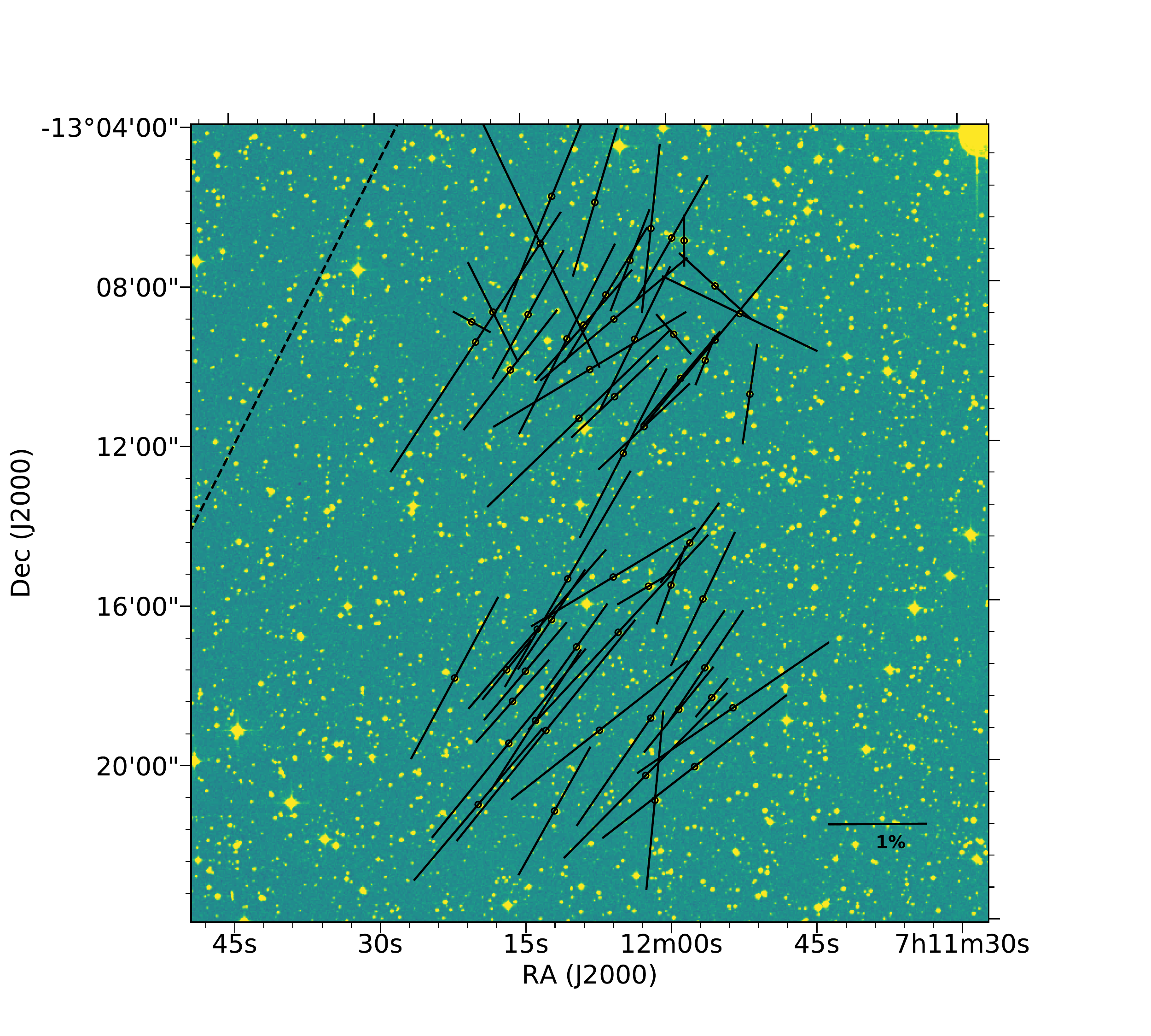}
\caption{I-Band} \label{fig:dss_F_I}  
\end{subfigure}

\caption{The polarization vectors in $B$, $V$, $R$, and $I$ bands are overplotted on 20\arcmin$\times$20\arcmin\, DSS $R$-band images of nearby field region of the cluster NGC 2345. The length of the vector shows the degree of polarization and their tilt denotes the position angle. The reference length for 1\% polarization is shown at the bottom right. The dotted line denotes the orientation of the projection of Galactic Parallel at the region.}
\label{fig:dss_field}
\end{figure*}

\twocolumn
\subsection{Degree of Polarization and Polarization Position Angle} \label{pth}

The polarization vectors in all four bands are overplotted on the DSS\footnote{http://archive.eso.org/dss/dss}-2-red survey images of the cluster and field regions  in Figures \ref{fig:dss} and \ref{fig:dss_field}, respectively. The length of the vector is proportional to the degree of polarization and their tilt denotes the position angle which is measured from north increasing towards the east. The reference vector for 1\% polarization is drawn at the bottom right. The member stars of the cluster NGC 2345 are marked by the filled red circles. The dotted line in figures is the projection of orientation of Galactic parallel (GP) at both regions. The majority of polarization vectors are found to be nearly parallel to the GP in both regions. There are a few stars in both regions that are showing a larger deviation from GP. 

\begin{figure}
\centering 
\begin{subfigure} {\columnwidth} 
\centering 
\includegraphics[width=\columnwidth]{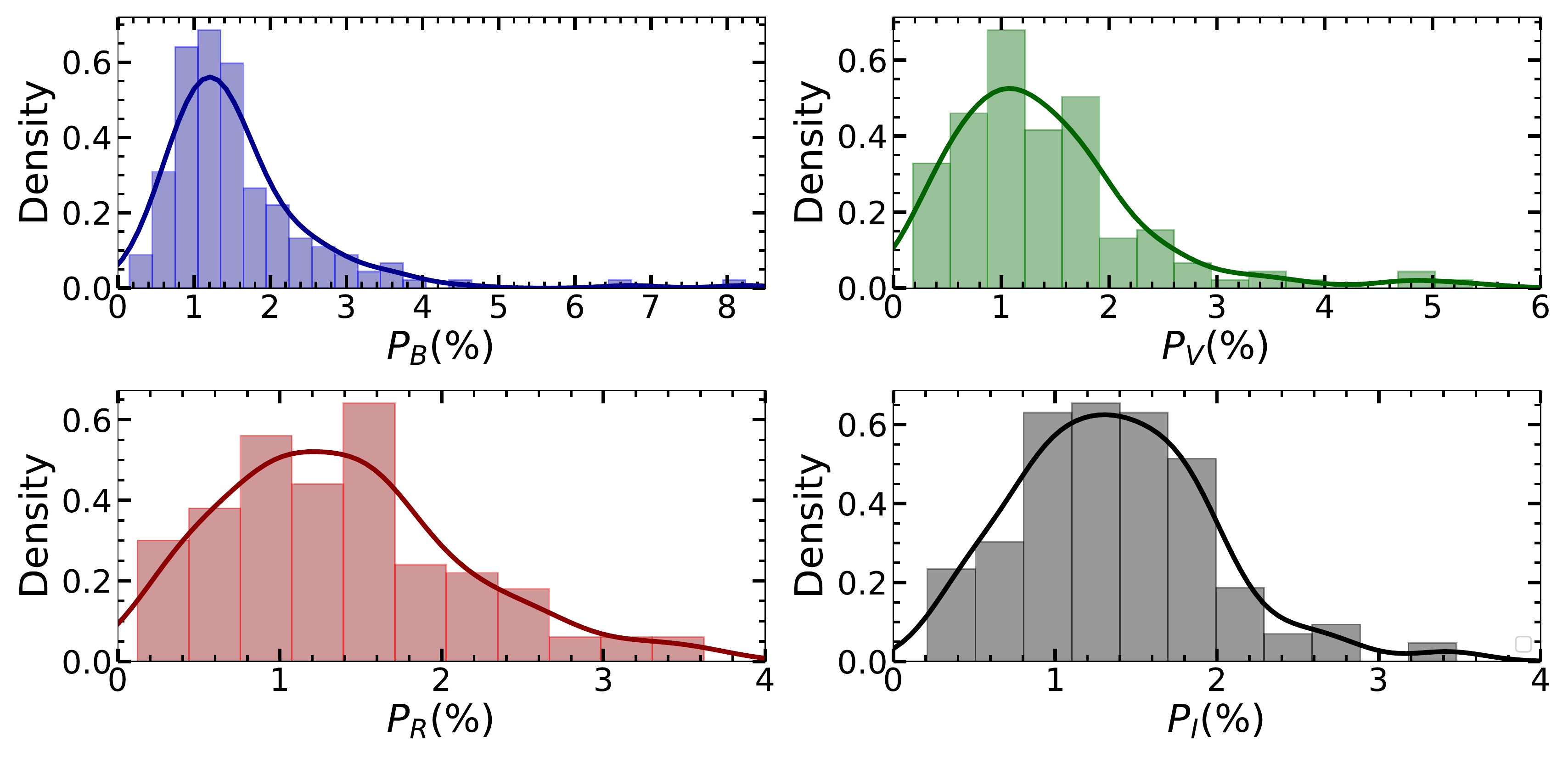}
\caption{The distribution of $P$ among observed stars.} \label{fig:phist}  
\end{subfigure}
\begin{subfigure} {\columnwidth} 
\centering 
\includegraphics[width=\columnwidth]{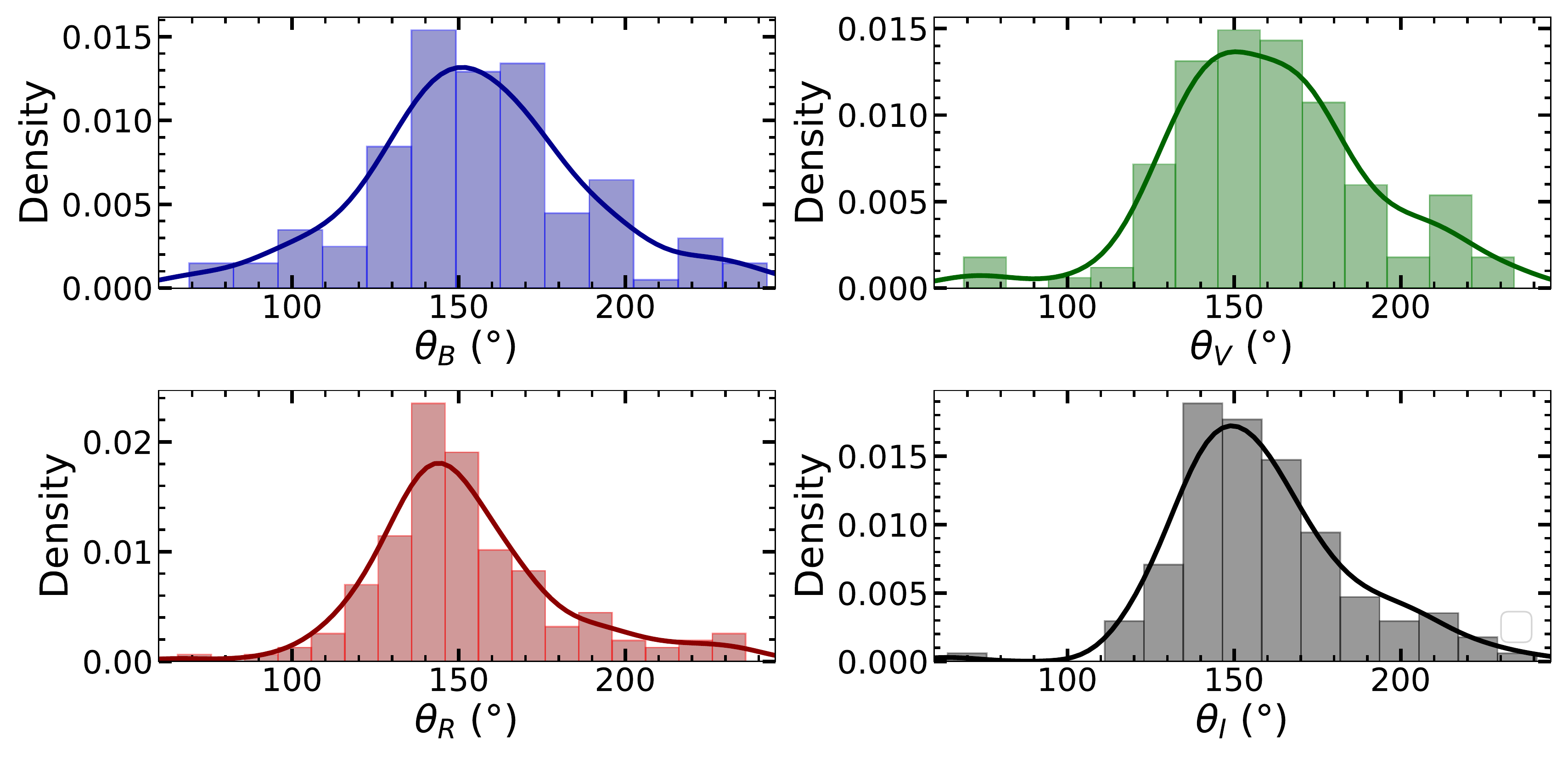}
\caption{The distribution of $\theta$.} \label{fig:thist} 
\end{subfigure}
\caption{(a, b) The distribution of $P$ and $\theta$ of observed stars. The blue, green, red, and black color shows the distributions in $B$, $V$, $R$, and $I$ bands, respectively. The Gaussian kernel density estimation is shown by the blue, green, red, and black curves for $B$, $V$, $R$, and $I$ bands, respectively.}
\label{fig:pthe_distri}
\end{figure}

In order to plot the distribution of $P$ and $\theta$, we have used the Gaussian kernel density estimation \citep[][]{scott1979optimal,2020zndo....592845W,Waskom2021}. The density distributions  along with the normalized histogram are shown in Figures \ref{fig:phist} and \ref{fig:thist} for $P$ and $\theta$ in $B$, $V$, $R$, and $I$-bands. As position angles indicate the orientation, hence circular statistics are considered for showing the distribution of position angles. The majority of stars have $P$ in the range of 0.5 $-$ 2\% and the histogram reveals that there is no such difference between the $\theta$ and the angle for GP ($\sim$ 153\degr\,), however, there is a dispersion in the value of position angle in all four bands.

\begin{figure}
\centering
\includegraphics[width=\columnwidth]{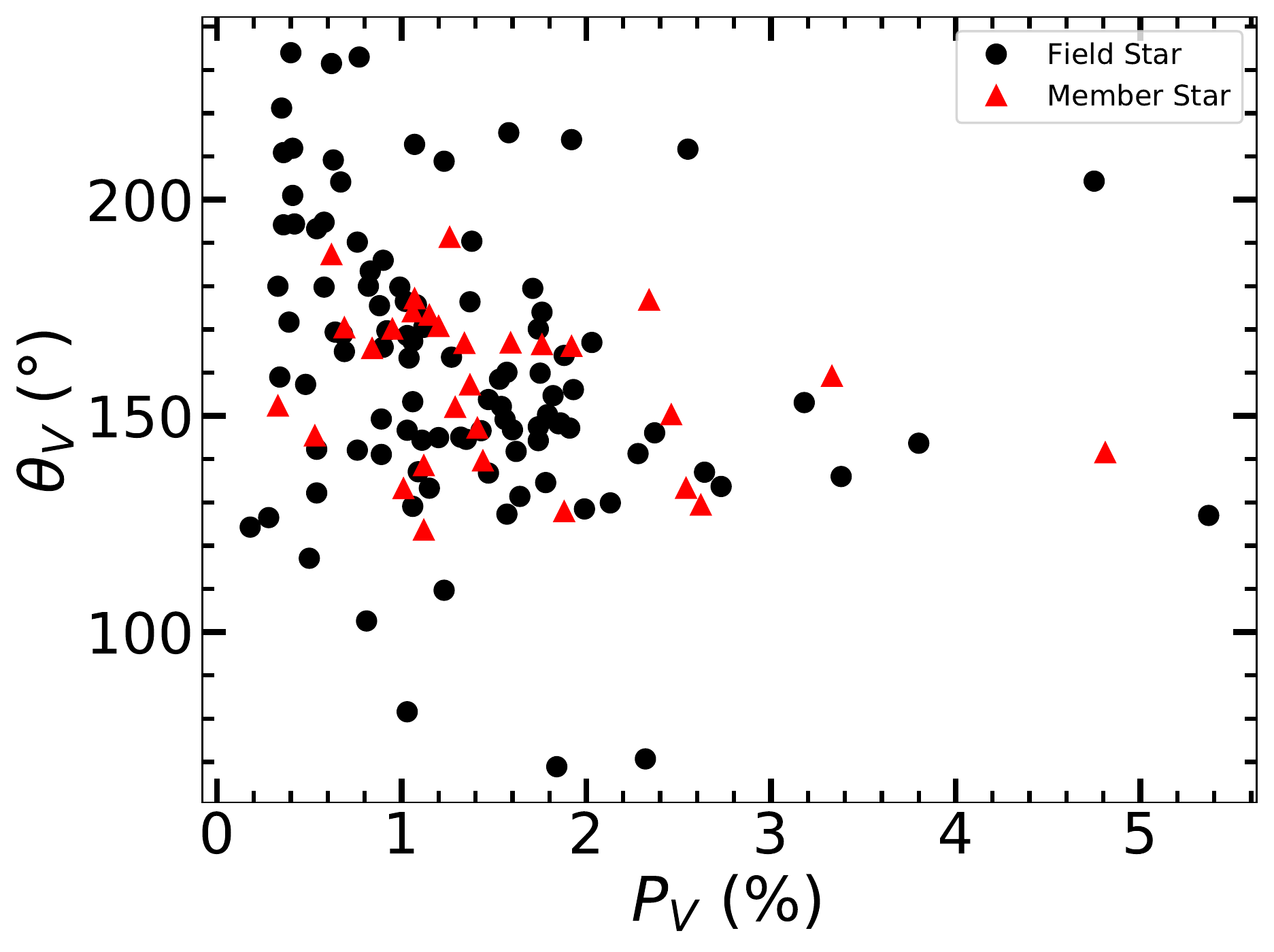}
\caption{The plot of $P_{V}$ versus $\theta_{V}$. Member stars of the cluster NGC 2345 are shown by the red-filled triangle and all other observed stars in the study are shown by the black-filled circle.}
\label{fig:pt_v}
\end{figure}

The plot between $P_{V}$ and $\theta_{V}$ for all observed stars is shown in Figure \ref{fig:pt_v} in which members of NGC 2345 and field stars are shown by red triangles and black circles, respectively. For the lower polarization, there is a large dispersion in the position angles. As $P_{V}$ increases then the spread in $\theta_{V}$ reduces i.e. the angle is more likely to be distributed near GP. The spread in angles is found to be more for field stars than the cluster member stars.

\subsection{Degree of Polarization, Position Angle, and Extinction as a function of Distance} \label{heiles}
The foreground dust concentration towards the cluster can be inferred by exploring the polarization and/or extinction versus distance plot. The polarization is supposed to increase with the distance if dust grains are uniformly aligned. However, in some past studies, the polarization was found to rise suddenly at some distances along the line of sight of the cluster and is inferred as the presence of dust layers at those distances \citep[e.g. see][]{2007A&A...462..621V,2008MNRAS.388..105M,2010MNRAS.403.1577M,2011MNRAS.411.1418E,2012MNRAS.419.2587E}. \citet{2010MNRAS.403.2041V,2018RMxAA..54..293V} have also shown the presence of dust layer based on the jumps in extinction but \citet{2018RMxAA..54..293V} did not found any rise in the polarization value at that distance.

\begin{figure}
\centering 
\begin{subfigure} {\columnwidth} 
\centering 
\includegraphics[width=\columnwidth]{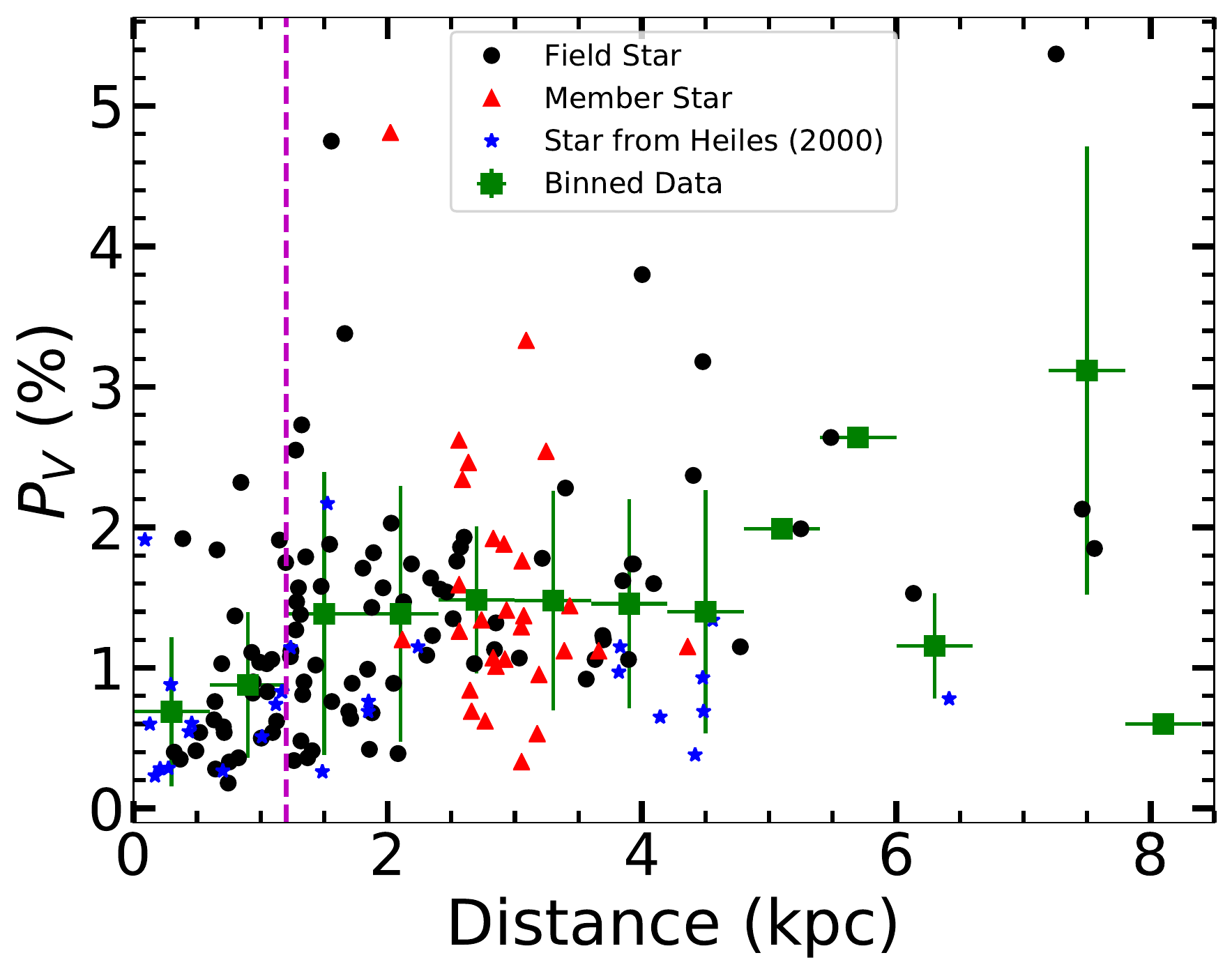}
\caption{} \label{fig:dist_p}  
\end{subfigure}
\begin{subfigure} {\columnwidth} 
\centering 
\includegraphics[width=\columnwidth]{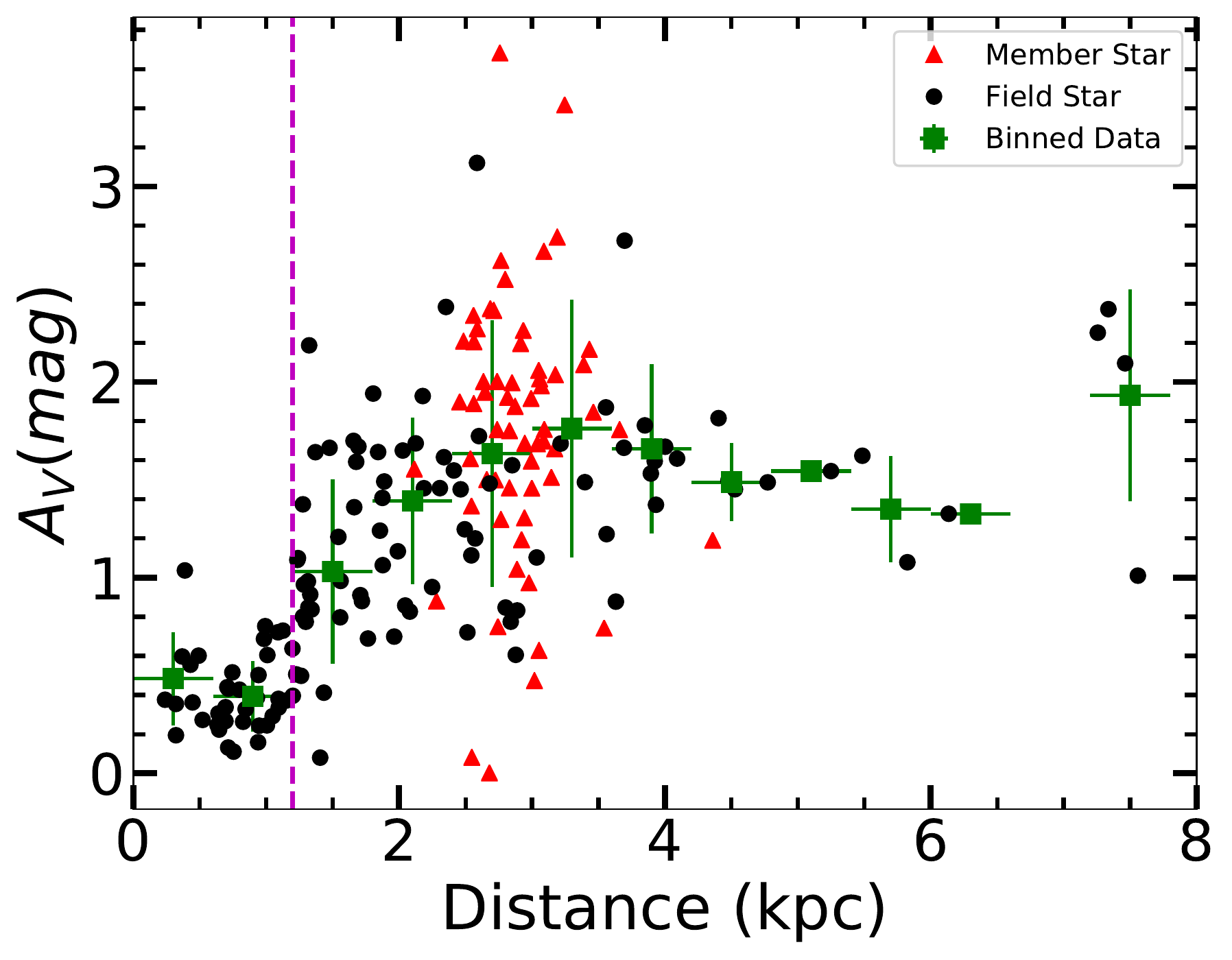}
\caption{} \label{fig:dist_av}  
\end{subfigure}
\begin{subfigure} {\columnwidth} 
\centering 
\includegraphics[width=\columnwidth]{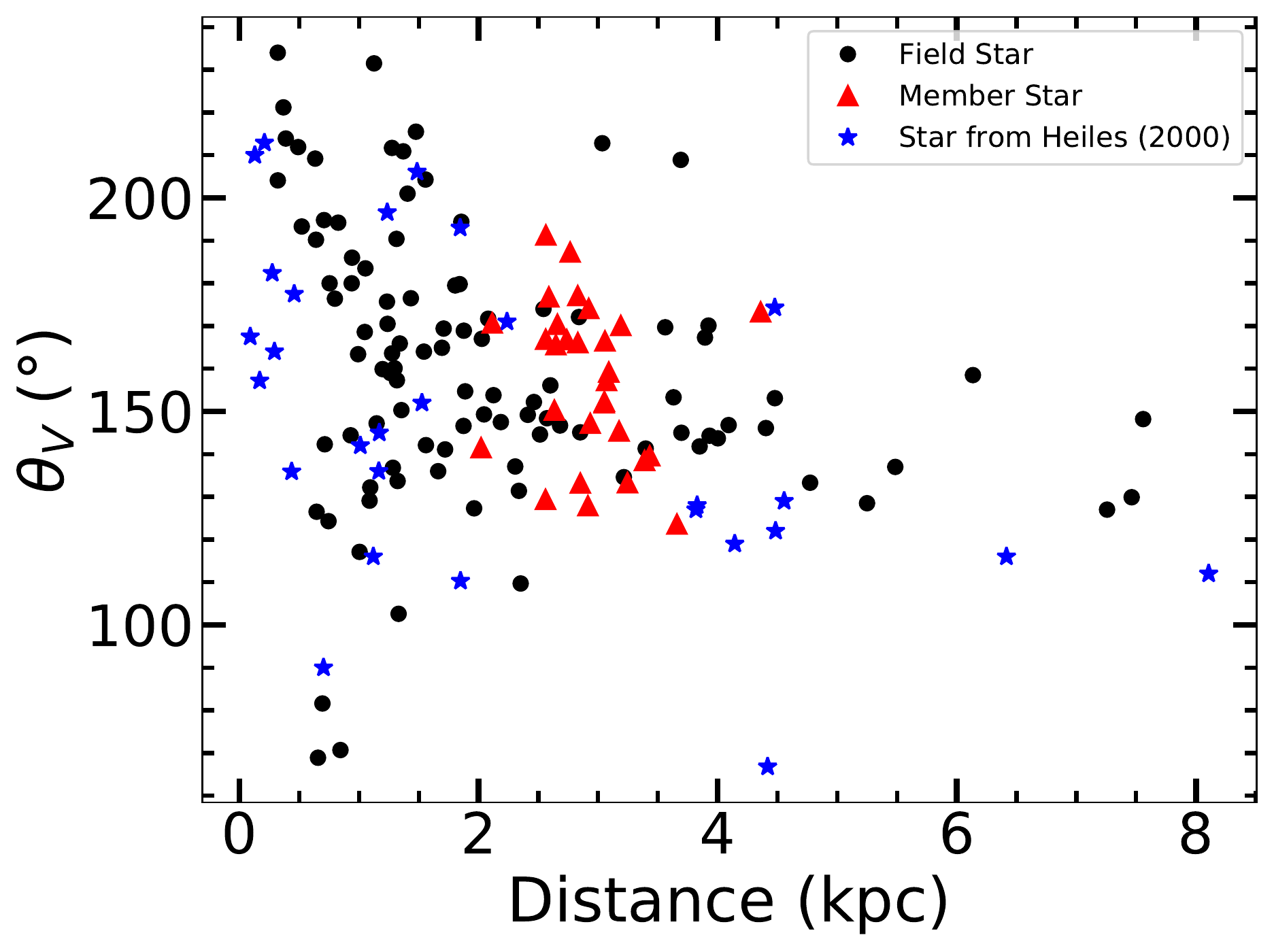}
\caption{} \label{fig:dist_th}  
\end{subfigure}
\caption{(a) Variation of degree of polarization in $V$-band with distance. The data has been binned for 0.6 kpc distance and shown by green-filled squares and error denotes the standard deviation. The magenta dotted vertical line at 1.2 kpc denotes the increment in polarization. (b) Variation of $A_{V}$ with distance. Red triangles are marking members. And binned data is shown for the 0.6 kpc binning scheme. (c) $\theta_{V}$ versus distance. Red triangles are for members, blue asterisks are stars taken from \citet{2000AJ....119..923H} and black circles are for all other observed stars.}
\label{fig:dist_pth}
\end{figure}

To explore the polarization properties of dust grains at different distances, we have plotted $P_{V}$, $\theta_{V}$, and extinction ($A_{V}$) as a function of distance in Figure \ref{fig:dist_pth}. The values of $A_{V}$ of stars are taken from \citet{2019A&A...628A..94A}. We have also taken polarization data from the catalogue of \citet{2000AJ....119..923H} within a 5\degr\, radius of the cluster centre. A total of 43 stars were found in this catalogue, out of which the polarization values for 15 stars were within 2$\sigma$ level and are not included in our analysis. Distances for stars were derived from the \textit{Gaia} DR2 parallaxes \citep[][]{2016A&A...595A...1G,2018A&A...616A...1G}, for stars whose parallax values are within 2$\sigma$ level were not considered for the analysis. There is an indication of a small increase in the degree of polarization at a distance of $\approx$ 1.2 kpc. We have also binned the data for distance 0.6 kpc and the average value of that bin is shown by green-filled squares in which the error bars indicate the standard deviation of that bin. After the distance of 1.2 kpc, the binned data is also showing a nearly constant trend. In addition to this and as seen in Figure \ref{fig:dist_av}, a rise in the extinction from the distance $\sim$ 1.2 kpc to a distance of a cluster is noticed. We have marked the increment position in polarization and extinction at 1.2 kpc with the vertical magenta dotted line. 

To get the enhancement in extinction after the distance 1.2 kpc, we have fitted a Gaussian curve to the distribution of $A_{V}$ for $d$ $\leq$ 1.2 and $d$ $>$ 1.2 kpc, which reveals an average increase of $\sim$ 1.15 mag in extinction after the distance 1.2 kpc. From Figure \ref{fig:dist_th}, it can be seen that, stars located before the distance $\sim$1.2 kpc, generally show more dispersion in position angle, whereas after $\sim$ 1.2 kpc, the position angle is distributed towards the GP direction with small spread. It can be clearly seen that as the distance increases after 1.2 kpc, the $\theta_{V}$ tends to decrease. The increment in the polarization and extinction after the distance $\sim$1.2 kpc towards the direction of the cluster NGC 2345 could be due to the enhancement and/or change in the distribution of dust grains.

\subsection{Wavelength-dependence Polarization} \label{ser}

The wavelength dependence of ISM polarization follows the relation \citep{1973IAUS...52..145S,1975ApJ...196..261S}, 
\begin{equation}
  P_\lambda = P_{\rm max} \exp[-1.15 \ln^2(\lambda_{\rm max}/\lambda)] 
  \label{eq:ser}
\end{equation}

where $P_\lambda$ is the polarization at wavelength $\lambda$, $P_{\rm max}$ is maximum polarization and $\lambda_{\rm max}$ is the wavelength corresponding to $P_{\rm max}$. The $\lambda_{\rm max}$ is a function of optical properties and characteristic particle size distribution of aligned grains. The relation provides the observed variation of ISM polarization with a wavelength between the range of 0.36 - 1 $\mu m $ \citep{1982AJ.....87..695W,1992ApJ...386..562W,2012MNRAS.419.2587E,2013ApJ...764..172P}. Using the above relation, we have estimated the parameters $P_{\rm max}$ and $\lambda_{\rm max}$ of those stars for which the data is available at least in three bands. We have computed the parameter $\sigma_{1}$ (unit weight error of fit), which quantifies the departure of data from the relation \ref{eq:ser}. The polarization is considered intrinsic in origin if $\sigma_{1}$ $>$ 1.6 \citep{1999AJ....117.2882W,2007A&A...462..621V,2008MNRAS.388..105M,2013ApJ...764..172P,2020AJ....160..256S}. Even if the relation \ref{eq:ser} fits well to the data, it is not necessary that the polarization is due to the ISM. There may be chances of another origin of polarization e.g. circumstellar shells follow a different wavelength dependency on polarization than the ISM one, which could resemble the interstellar law in a limited range of wavelength \citep{1998AJ....116..266O}. Thus stars with  the value of $\lambda_{\rm max}$ considerably smaller or larger than the average value for general ISM (0.55 $\mu m$), may be considered as intrinsically polarized \citep{1975ApJ...196..261S, 1998AJ....116..266O,2007A&A...462..621V,2010MNRAS.403.1577M,2020AJ....159...99S,2020AJ....160..256S}. 

\begin{table*}
\centering
\large
\setlength{\tabcolsep}{12pt}
\caption{The value of $P_{\rm max}$, $\lambda_{\rm max}$, and $\sigma_{1}$ as obtained after best fit of Serkowski relation.}
\label{tab:ser}
\begin{tabular}{|cccc|ccccc}
\hline
ID & $P_{\rm max}$  & $\lambda_{\rm max}$  & $\sigma_{1}$ & & ID & $P_{\rm max}$  & $\lambda_{\rm max}$  & $\sigma_{1}$ \\
 & (\%) & ($\mu m$) & & & & (\%) & ($\mu m$) & \\
\hline
1     &   1.88$\pm$0.03  &  0.84$\pm$0.02  &  0.1    & & 	90    &   1.20$\pm$0.09  &  0.62$\pm$0.12  &  0.7 \\
4     &   1.67$\pm$0.12  &  0.53$\pm$0.07  &  0.5    & & 	92    &   1.05$\pm$0.17  &  0.66$\pm$0.27  &  0.9 \\
5     &   1.73$\pm$0.03  &  0.74$\pm$0.04  &  0.2    & & 	93    &   4.57$\pm$0.70  &  0.16$\pm$0.01  &  0.3 \\
6     &   2.79$\pm$0.08  &  0.69$\pm$0.04  &  1.5    & & 	98    &   2.03$\pm$0.05  &  0.59$\pm$0.03  &  0.9 \\
9     &   1.87$\pm$0.02  &  0.56$\pm$0.01  &  0.2    & & 	136   &   1.45$\pm$0.02  &  0.78$\pm$0.04  &  0.2 \\
10    &   1.97$\pm$0.07  &  0.44$\pm$0.02  &  0.6    & & 	137   &   1.88$\pm$0.11  &  0.57$\pm$0.08  &  1.1 \\
11    &   2.88$\pm$0.09  &  0.40$\pm$0.01  &  0.2    & & 	138   &   0.96$\pm$0.31  &  0.44$\pm$0.12  &  1.0 \\
12    &   2.56$\pm$0.62  &  0.31$\pm$0.09  &  1.2    & & 	139   &   1.17$\pm$0.20  &  0.73$\pm$0.20  &  1.1 \\
13    &   1.56$\pm$0.02  &  0.64$\pm$0.01  &  0.1    &  &	140   &   1.31$\pm$0.23  &  0.45$\pm$0.08  &  0.8 \\
14    &   1.11$\pm$0.09  &  0.69$\pm$0.12  &  0.4    & & 	141   &   2.91$\pm$0.63  &  0.36$\pm$0.07  &  0.9 \\
16    &   0.98$\pm$0.28  &  0.41$\pm$0.13  &  1.5    & & 	142   &   2.44$\pm$0.08  &  0.46$\pm$0.03  &  0.3 \\
17    &   2.74$\pm$0.60  &  1.25$\pm$0.12  &  0.7    & & 	143   &   1.77$\pm$0.77  &  0.32$\pm$0.11  &  0.8 \\
18    &   3.07$\pm$0.48  &  0.29$\pm$0.02  &  0.6    & & 	144   &   1.64$\pm$0.02  &  0.53$\pm$0.02  &  0.2 \\
24    &   1.80$\pm$0.05  &  0.57$\pm$0.03  &  0.6    &  &	145   &   2.61$\pm$0.07  &  0.55$\pm$0.02  &  0.2 \\
25    &   2.04$\pm$0.08  &  0.68$\pm$0.07  &  0.9    & & 	146   &   1.49$\pm$0.30  &  0.88$\pm$0.19  &  0.5 \\
26    &   1.85$\pm$0.56  &  0.35$\pm$0.07  &  0.9    & & 	147   &   1.90$\pm$0.15  &  0.64$\pm$0.16  &  0.8 \\
28    &   1.16$\pm$0.15  &  0.59$\pm$0.11  &  1.0    & & 	149   &   2.47$\pm$0.33  &  0.24$\pm$0.02  &  0.6 \\
29    &   2.21$\pm$0.35  &  0.53$\pm$0.16  &  1.1    & & 	150   &   3.95$\pm$0.17  &  0.37$\pm$0.03  &  1.3 \\
32    &   1.20$\pm$0.15  &  0.62$\pm$0.14  &  0.6    & & 	151   &   6.35$\pm$0.42  &  0.28$\pm$0.01  &  0.1 \\
34    &   2.45$\pm$0.39  &  0.55$\pm$0.10  &  2.5    & & 	153   &   0.81$\pm$0.19  &  0.49$\pm$0.12  &  0.6 \\
35    &   1.58$\pm$0.09  &  0.61$\pm$0.10  &  0.8    & & 	154   &   0.60$\pm$0.10  &  0.72$\pm$0.19  &  0.4 \\
36    &   1.72$\pm$0.05  &  0.53$\pm$0.04  &  0.2    & & 	155   &   1.90$\pm$0.12  &  0.37$\pm$0.02  &  0.2 \\
37    &   1.68$\pm$0.31  &  0.82$\pm$0.13  &  0.8    & & 	157   &   1.08$\pm$0.19  &  0.58$\pm$0.23  &  0.9 \\
38    &   1.68$\pm$0.31  &  0.68$\pm$0.21  &  1.0    & & 	158   &   1.28$\pm$0.45  &  0.85$\pm$0.31  &  1.0 \\
39    &   1.07$\pm$0.03  &  0.60$\pm$0.08  &  0.4    & & 	159   &   1.68$\pm$0.10  &  0.54$\pm$0.07  &  1.5 \\
40    &   2.71$\pm$0.75  &  0.88$\pm$0.28  &  2.5    &  &	162   &   2.04$\pm$0.57  &  0.92$\pm$0.33  &  0.8 \\
42    &   1.41$\pm$0.02  &  0.68$\pm$0.02  &  0.2    & & 	163   &   2.36$\pm$0.51  &  0.56$\pm$0.16  &  1.3 \\
44    &   1.62$\pm$0.08  &  0.85$\pm$0.04  &  0.3    & & 	164   &   1.09$\pm$0.02  &  0.52$\pm$0.02  &  0.2 \\
46    &   1.77$\pm$0.13  &  0.57$\pm$0.07  &  1.0    & & 	166   &   0.39$\pm$0.14  &  0.82$\pm$0.33  &  0.5 \\
47    &   0.54$\pm$0.26  &  0.33$\pm$0.12  &  0.4    & & 	167   &   2.80$\pm$0.15  &  1.11$\pm$0.05  &  0.1 \\
48    &   3.17$\pm$1.32  &  0.35$\pm$0.15  &  1.2    & & 	168   &   2.47$\pm$0.21  &  0.33$\pm$0.02  &  0.2 \\
49    &   3.94$\pm$0.76  &  0.51$\pm$0.16  &  1.2    & & 	169   &   1.93$\pm$0.15  &  0.48$\pm$0.12  &  0.9 \\
51    &   1.65$\pm$0.35  &  0.46$\pm$0.17  &  1.3    & & 	170   &   1.59$\pm$0.04  &  0.65$\pm$0.21  &  1.0 \\
53    &   0.86$\pm$0.25  &  0.38$\pm$0.06  &  0.3    & & 	172   &   2.65$\pm$0.50  &  1.50$\pm$0.19  &  0.3 \\
63    &   1.13$\pm$0.02  &  0.75$\pm$0.03  &  0.9    & & 	174   &   2.71$\pm$0.93  &  0.48$\pm$0.16  &  1.2 \\
64    &   1.44$\pm$0.20  &  0.83$\pm$0.12  &  0.3    & & 	175   &   2.41$\pm$0.22  &  0.56$\pm$0.08  &  0.7 \\
66    &   2.48$\pm$0.27  &  0.36$\pm$0.03  &  0.7    & & 	176   &   0.52$\pm$0.08  &  0.36$\pm$0.05  &  0.6 \\
67    &   1.91$\pm$0.07  &  0.64$\pm$0.04  &  0.5    & & 	177   &   1.90$\pm$0.58  &  1.54$\pm$0.31  &  0.3 \\
68    &   2.67$\pm$0.04  &  0.69$\pm$0.03  &  0.4    & & 	179   &   3.83$\pm$1.10  &  0.97$\pm$0.32  &  2.8 \\
69    &   1.69$\pm$0.06  &  0.66$\pm$0.03  &  0.6    &&  	180   &   1.62$\pm$0.16  &  0.65$\pm$0.16  &  0.7 \\
70    &   1.00$\pm$0.25  &  0.43$\pm$0.19  &  0.8    & & 	184   &   0.91$\pm$0.04  &  0.66$\pm$0.04  &  1.0 \\
71    &   1.75$\pm$0.17  &  0.75$\pm$0.10  &  1.5    & & 	185   &   1.36$\pm$0.39  &  0.81$\pm$0.34  &  1.8 \\
72    &   1.65$\pm$0.04  &  0.70$\pm$0.05  &  0.3    & & 	186   &   1.43$\pm$0.01  &  0.55$\pm$0.02  &  0.2 \\
74    &   2.84$\pm$0.84  &  0.81$\pm$0.23  &  1.9    & & 	187   &   0.35$\pm$0.06  &  0.48$\pm$0.17  &  0.3 \\
77    &   3.51$\pm$0.03  &  0.74$\pm$0.01  &  0.3    & & 	189   &   1.93$\pm$0.23  &  0.81$\pm$0.15  &  0.7 \\
82    &   0.76$\pm$0.12  &  0.36$\pm$0.06  &  0.4    & & 	190   &   1.16$\pm$0.47  &  0.36$\pm$0.11  &  0.6 \\
85    &   0.50$\pm$0.04  &  0.37$\pm$0.03  &  0.1    & & 	192   &   1.06$\pm$0.08  &  0.43$\pm$0.12  &  0.9 \\
86    &   0.95$\pm$0.03  &  0.59$\pm$0.07  &  0.6    & & 	193   &   1.92$\pm$0.55  &  0.36$\pm$0.08  &  1.2 \\
87    &   1.57$\pm$0.24  &  0.60$\pm$0.21  &  1.0    & & 	195   &   4.34$\pm$0.41  &  0.29$\pm$0.01  &  0.1 \\
89    &   1.15$\pm$0.11  &  0.72$\pm$0.09  &  0.5    & & 	196   &   1.60$\pm$0.26  &  0.47$\pm$0.09  &  0.6 \\
\hline
\end{tabular}
\end{table*}

\begin{figure}
\centering 
\includegraphics[width=\columnwidth]{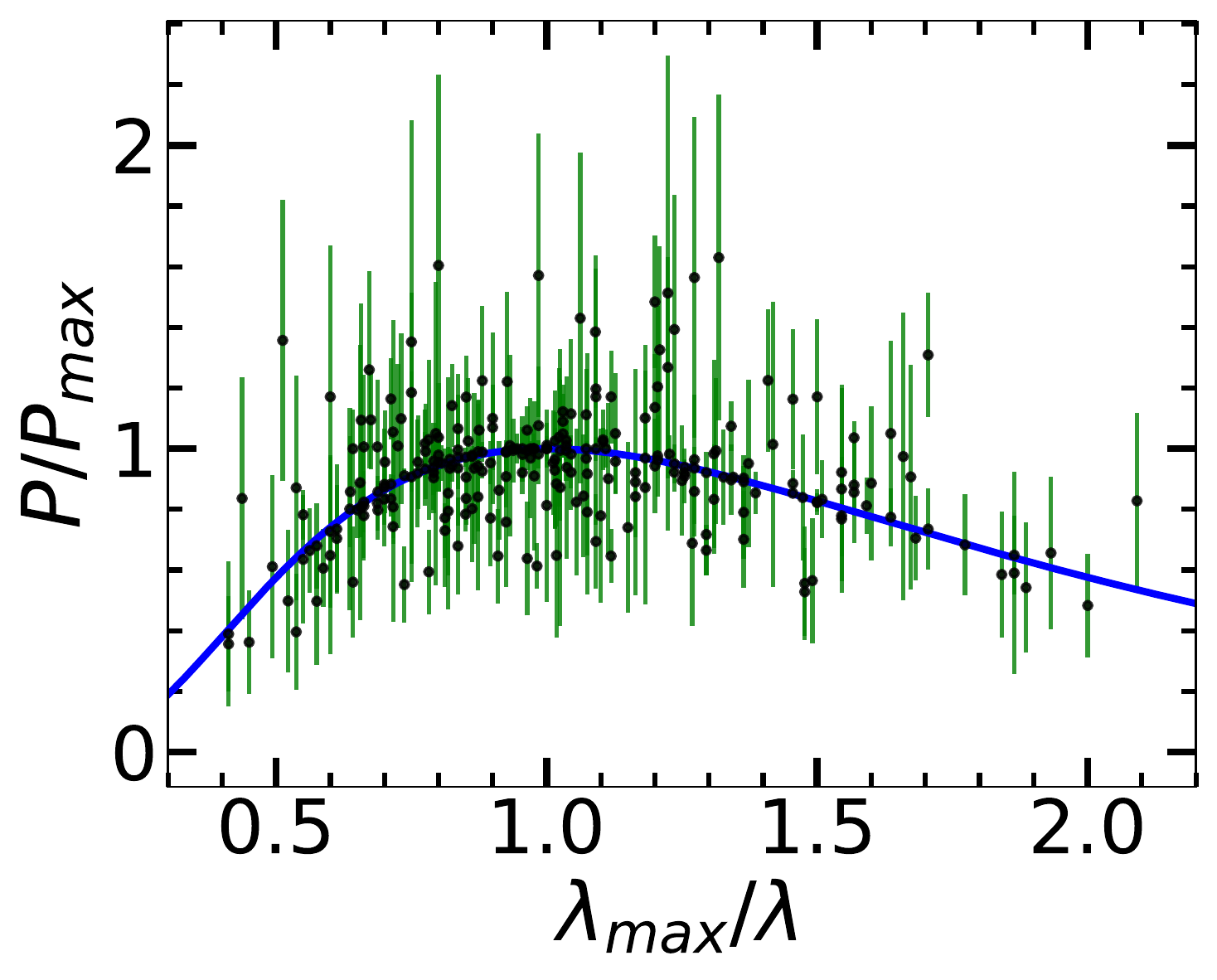}
\caption{Normalized wavelength-dependence-polarization plot for observed stars, intrinsically polarized stars are excluded from the plot. The blue curve shows the Serkowski relation.}
\label{fig:norm_ser}
\end{figure}

The derived value of $P_{\rm max}$, $\lambda_{\rm max}$, and $\sigma_{1}$ for 100 stars are given in Table \ref{tab:ser}, where stars' IDs are same as in Table \ref{tab:pthe}. In this table, we have only given those stars for which the values of $P_{\rm max}$ and/or $\lambda_{\rm max}$ are determined above 2$\sigma$ level. Based on $\sigma_{1}$ criteria, we found that five stars (ID: 34, 40, 74, 179, and 185) have an intrinsic component of polarization.  For the ISM originated polarization the value of $\lambda_{\rm max}$ should be in the range of 0.45 $\mu m$ - 0.8 $\mu m$ \citep{1975ApJ...196..261S}. For 19 stars (ID: 11, 12, 18, 26, 53, 66, 82, 85, 93, 141, 143, 149, 150, 151, 155, 168, 176, 193, and 195), the value of $\lambda_{\rm max}$ is found to be significantly less than 0.45 $\mu m$, whereas for six stars (ID: 1, 17, 44, 167, 172, and 177), $\lambda_{\rm max}$ is found to have significantly higher value than 0.8 $\mu m$. Thus, there are a total of 30 stars which may also have an intrinsic component of polarization. Two stars (ID 11 and 82) are reported as blue stars in \citet{1971PW&SO...1a...1S,1974A&AS...16...33M, 2019A&A...631A.124A} and there is a probability for the presence of significant component of intrinsic polarization in blue stars \citep{1969AJ.....74..528C,1973ApJ...184..173C,1976ApJ...208..253H,1978ApJS...38..229P}. Excluding all intrinsically polarized stars, the average value of $\lambda_{\rm max}$ and $P_{\rm max}$ were determined after Gaussian fitting to their distribution and found to be  0.58$\pm$0.13 $\mu m$ and 1.55$\pm$0.47 \%, respectively, where the error in each parameter is standard deviation. The plot between $\lambda_{\rm max}/\lambda$ and $P/P_{\rm max}$ is shown in Figure \ref{fig:norm_ser}, where the blue curve shows the Serkowski relation for general ISM. The value of $\lambda_{\rm max}$ is associated with the average size of ISM dust grains \citep{1968ApJ...154..115S,1973IAUS...52..145S} and estimated to be around 0.55 $\mu m$ for general ISM. This indicates that the average size distribution of dust grains towards the direction of NGC 2345 is similar to that of general ISM. 

\subsection{Polarizing Efficiency} \label{eff}
\begin{figure}
\centering 
\begin{subfigure} {\columnwidth} 
\centering 
\includegraphics[width=\columnwidth]{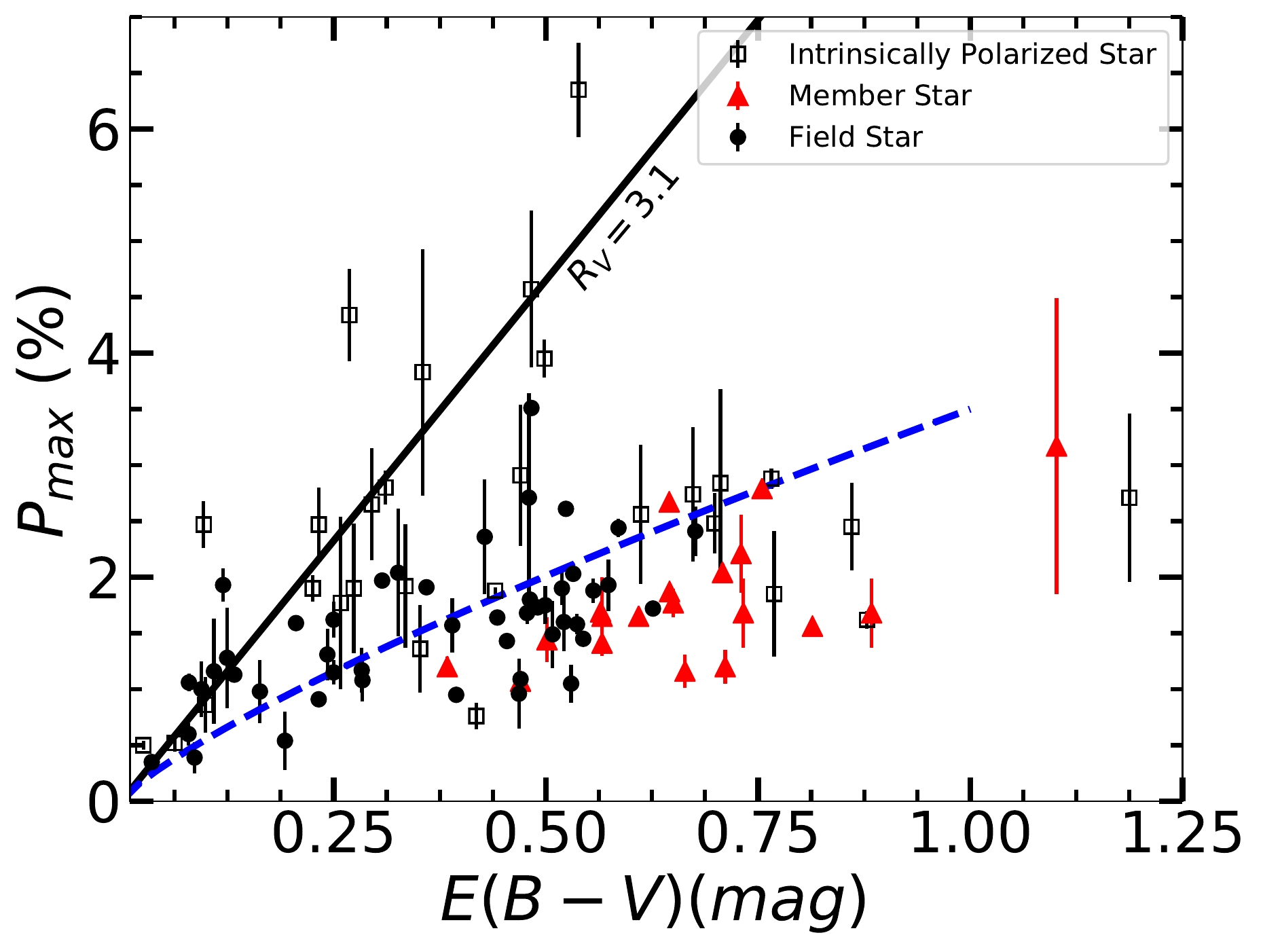}
\caption{} \label{fig:eff1}
 \end{subfigure}
\begin{subfigure} {\columnwidth} 
\centering 
\includegraphics[width=\columnwidth]{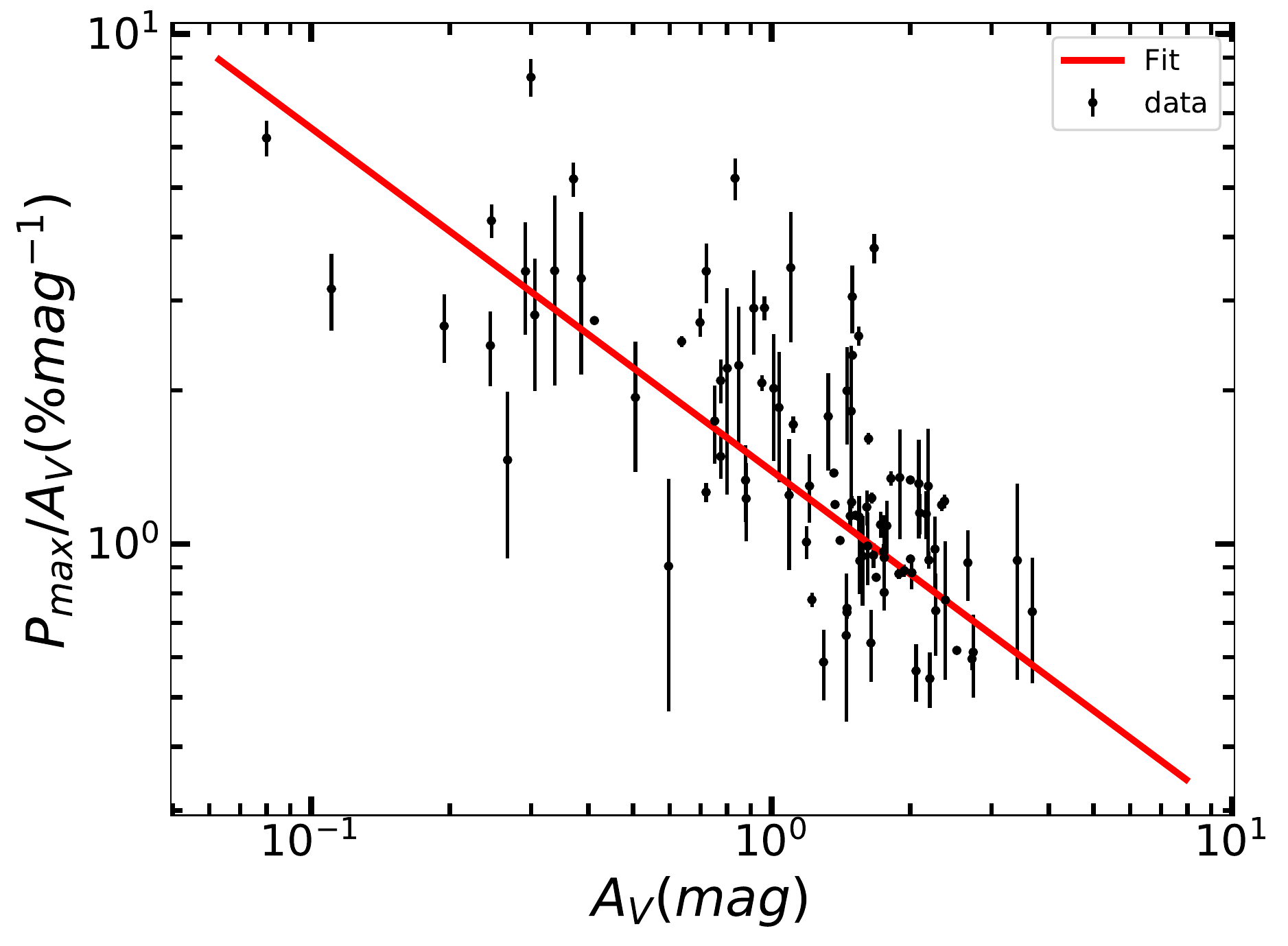}
\caption{} \label{fig:eff2}  
\end{subfigure}
\caption{The polarizing efficiency diagram. (a) The variation of $P_{\rm max}$ with reddening $E(B-V)$. The black line shows the line for maximum efficiency for $R_{V}$ =3.1, while the blue curve signifies average efficiency estimated for Galaxy. (b) $P_{\rm max}/A_{V}$ as a function of $A_{V}$. The red line shows the best-fit power law to the data in the log-log scale using the bootstrap sampling method.}
\label{fig:eff}
\end{figure}

The polarizing efficiency is a measure of the polarization produced by a given amount of extinction (i.e. the ratio of polarization to the reddening or extinction). Polarizing efficiency depends upon the magnetic field orientation, strength, and alignment efficiency \citep{2008JQSRT.109.1527V,2012A&A...541A..52V,2012JQSRT.113.2334V}. Thus different regions e.g. molecular clouds, diffuse ISM, and star-forming regions etc. of Galaxy show different behaviour of polarizing efficiency. Figure \ref{fig:eff} shows the polarizing efficiency plots for the observed region NGC 2345. For the diffused ISM, the maximum polarization for a given extinction can be limited by $P_{\rm max} \leq 3 R_{V} \times E(B-V)$ \citep{1956ApJS....2..389H,1973IAUS...52..145S,1975ApJ...196..261S}. The black straight line in Figure \ref{fig:eff1} corresponds to the maximum polarization with total-to-selective extinction ($R_{V}$) as 3.1. Polarizing efficiency for the majority of stars is found to be below the maximum efficiency for the region NGC 2345. There are only a few stars that show the polarizing efficiency more than the maximum value expected for the region which could be due to the intrinsic nature of polarization in these stars. The blue dotted curve in Figure \ref{fig:eff1} shows the average value of polarizing efficiency of the Galaxy which is estimated by using the relation  $P = 3.5 \times E(B-V)^{0.8}$ for $E(B-V) < $ 1.0 mag \citep{2002ApJ...564..762F}. It is seen from this figure the values of polarization for the majority of cluster members were found to be less than the average value for the Galaxy. The majority of cluster members are lying below the average efficiency curve (dashed blue curve). However, the majority of field stars lies in between the maximum estimated value and the average value of polarizing efficiency for the Galaxy. This analysis further indicates that polarization and polarizing efficiency in observed regions is similar to the general ISM. 

The alignment of grains in the region can be examined by the behaviour of  polarizing efficiency with extinction. A number of studies \citep[e.g.][]{1992ApJ...389..602J,1994MNRAS.268....1W,2001ApJ...547..872W,2008ApJ...674..304W,2014A&A...569L...1A,2015ARA&A..53..501A,2015MNRAS.448.1178H,2015AJ....149...31J,2017ApJ...849..157W,2018MNRAS.475.5535I,2019ApJ...873...87M,2021AJ....161..149S} have found reducing polarizing efficiency with extinction, which follows the power-law relation $P_{\rm max}/A_{V}=\beta (A_{V})^{\alpha}$ with power-law index, $\alpha$ close to -0.5. For the inner region of the clouds, the value of $\alpha$ near -1 denotes the total loss of grain alignment which occur for $A_{V}$ $\gtrsim$ 20 mag \citep{2015AJ....149...31J}.  We have also fitted the power law, which yields the relation $P_{\rm max}/A_{V} = (1.39\pm0.09) (A_{V})^{-0.67\pm0.09}$ for the region of the cluster NGC 2345. We used the bootstrap method \citep{10.1214/aos/1176344552,efron1994introduction} for estimating the value of the power-law index and corresponding error. In the bootstrap method, the sample statics is determined using random sampling with replacement, which increases the accuracy of sample estimates. The re-sampling is carried out 2000 times with a similar sample size as data points and replacement. The power-law fit is performed for each sample and the mean value of fitting parameters with the standard deviation is calculated. Figure \ref{fig:eff2} shows the variation of $P_{\rm max}/A_{V}$ with $A_{V}$ along with the best fit power law. Here, for better representation, the data and corresponding best-fit power-law are converted into the log-log scale.

\begin{figure}
\centering 
\begin{subfigure} {\columnwidth} 
\centering 
\includegraphics[width=\columnwidth]{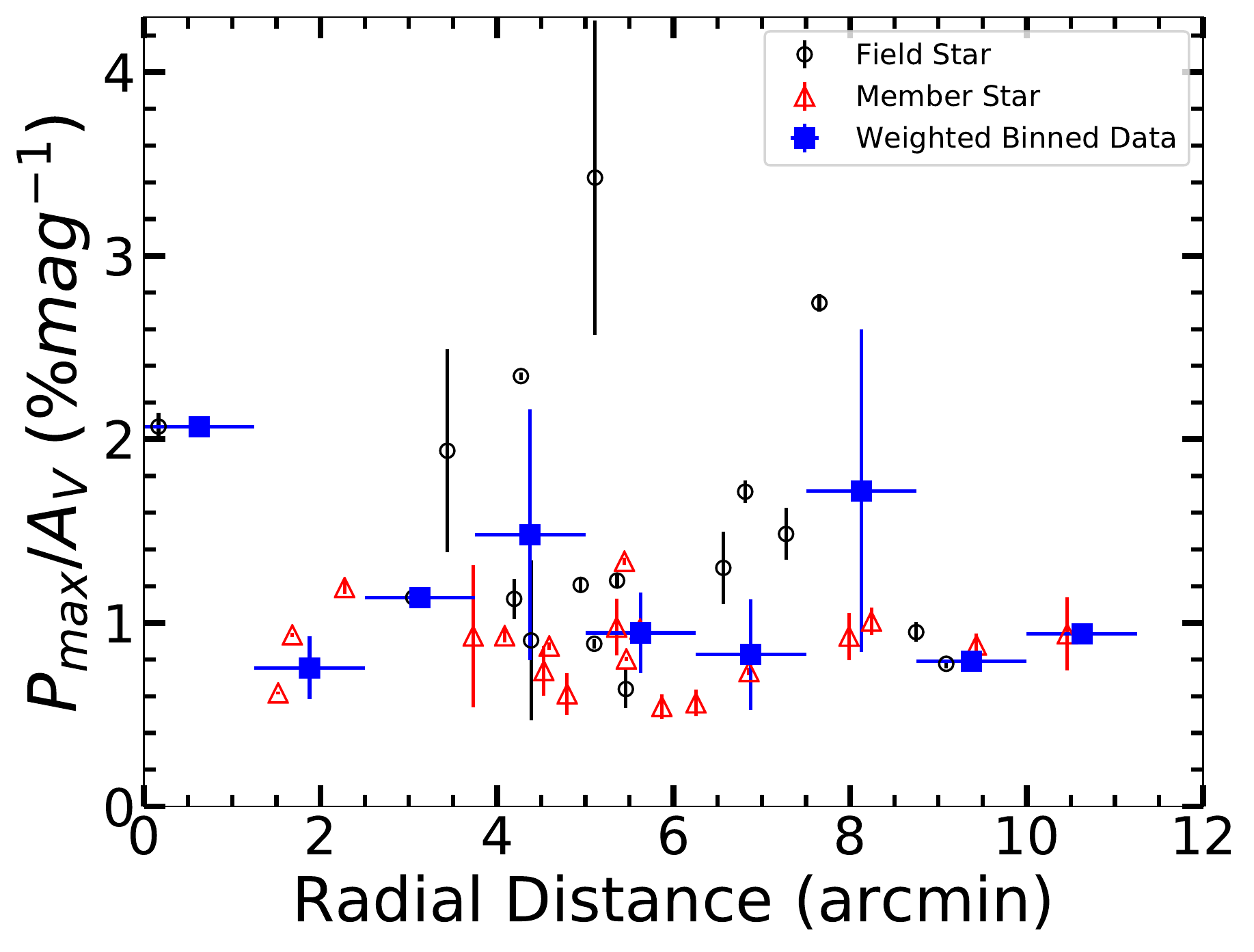}
\caption{} \label{fig:spatial_pmax_av}  
\end{subfigure}
\begin{subfigure} {\columnwidth} 
\centering 
\includegraphics[width=\columnwidth]{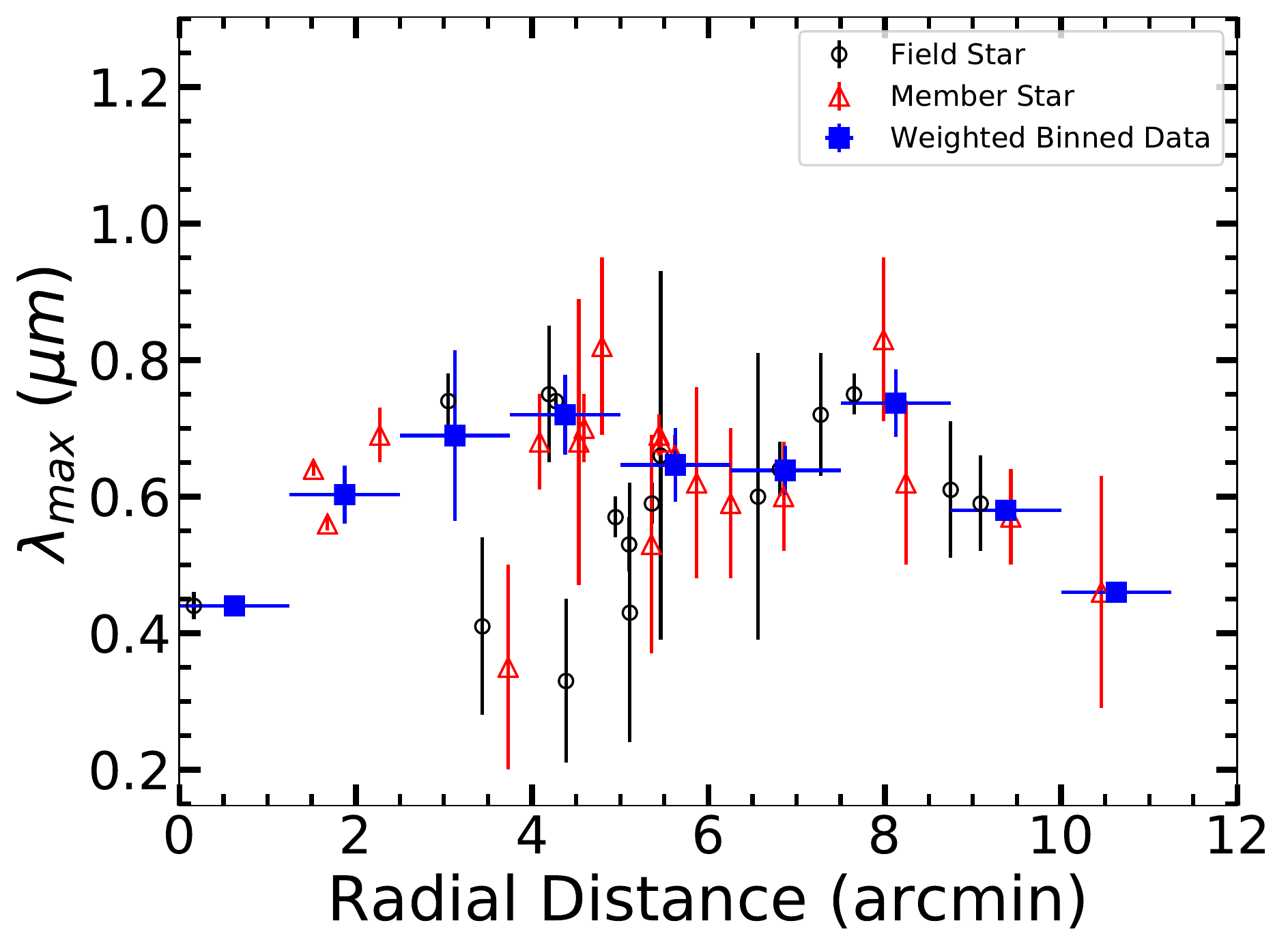}
\caption{} \label{fig:spatial_lmax1}  
\end{subfigure}
\caption{The distribution of $P_{\rm max}/A_{V}$ and  $\lambda_{\rm max}$ with the radial distance from the centre of the cluster NGC 2345. The data has been binned for $\sim$ 1\arcmin.25 along the radial direction, where each bin value is the weighted average of the parameter in that bin and error bars show the weighted standard deviation.}
\label{fig:spatial}
\end{figure}

The distribution of $P_{\rm max}/A_{V}$ and $\lambda_{\rm max}$ with the radial distance from the centre of cluster NGC 2345 are shown in Figure \ref{fig:spatial}. Intrinsically polarized stars are not included in this plot. The data has been binned for $\sim$ 1\arcmin.25 along the radial direction. The binned data is shown by blue squares. Each bin value is the weighted average of the parameter in that bin and the weighted standard deviation is shown by error bars. The $P_{\rm max}/A_{V}$ was found to be decreasing from centre to outwards of the cluster NGC 2345. A higher value of $P_{\rm max}/A_{V}$ is seen in the core region with respect to the coronal region of the cluster. Whereas an increasing trend in the value of $\lambda_{\rm max}$ was found from core region to outwards. 

\subsection{Wavelength of Maximum Polarization as function of Extinction and Polarizing Efficiency} \label{lmax_var}
\begin{figure*}
\centering 
\begin{subfigure} {\columnwidth} 
\centering 
\includegraphics[width=\columnwidth]{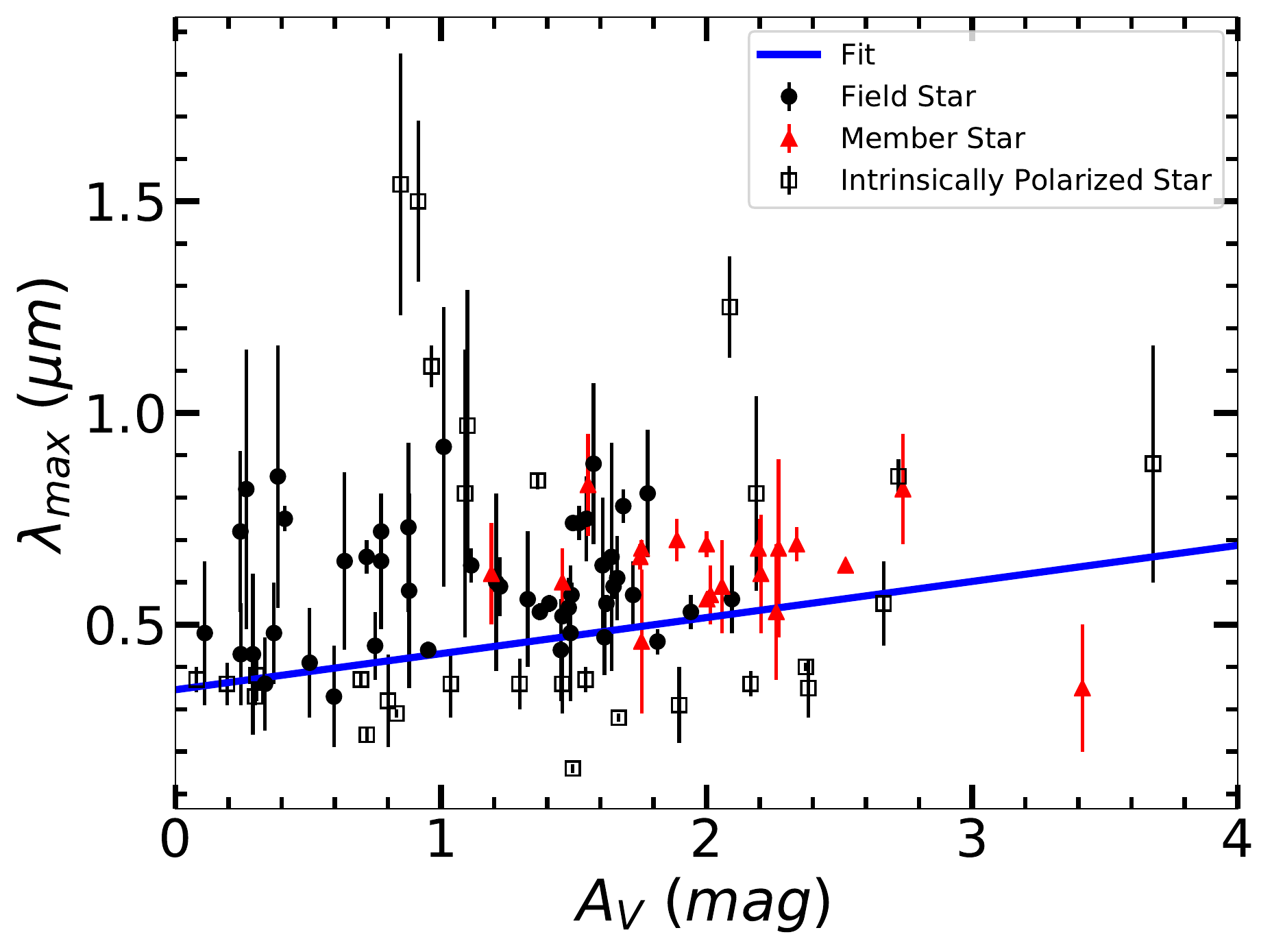}
\caption{} \label{fig:lmax_av}
\end{subfigure}
\begin{subfigure} {\columnwidth} 
\centering 
\includegraphics[width=\columnwidth]{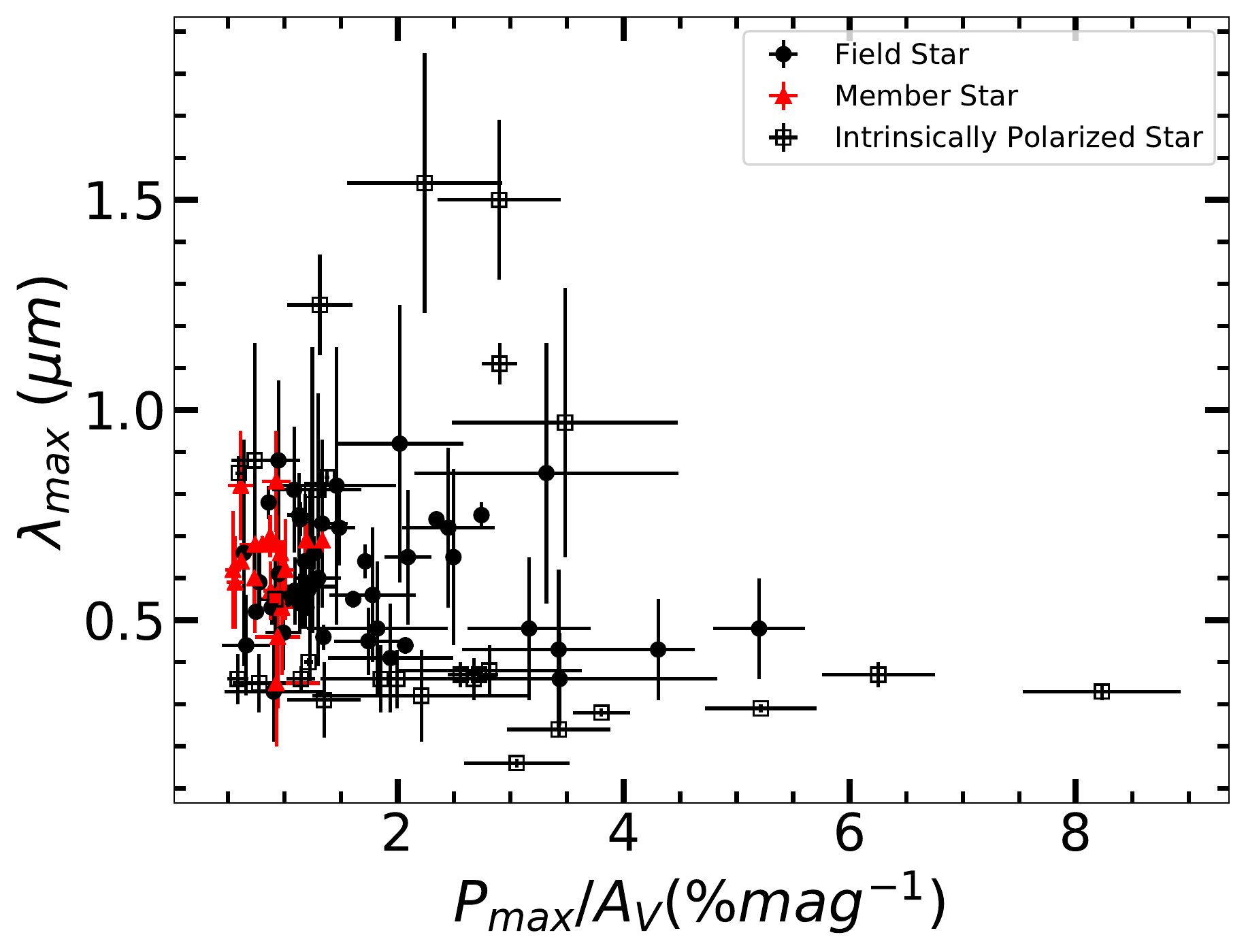}
\caption{} \label{fig:eff_lmax}  
\end{subfigure}
\caption{(a) Variation of $\lambda_{\rm max}$ with $A_{V}$. The data have been fitted with a straight line represented by the blue line. (b) The variation of $\lambda_{\rm max}$ with the polarizing efficiency for the observed region.}
\label{fig:lmax_var}
\end{figure*}

\citet{1978A&A....66...57W} had shown a relation between $R_{V}$ and $\lambda_{\rm max}$  as $R_{V} = 5.6 \times \lambda_{\rm max}$. However, in the later studies it was found that no clear trend of increasing $R_{V}$ with $\lambda_{\rm max}$, a clustering around the value of $\lambda_{\rm max}$ near 0.55 $\mu m$ in some studies \citep[e.g.][]{1988ApJ...327..911C,1994MNRAS.268....1W,2001ApJ...547..872W,2007ApJ...665..369A} was noticed. The poor correlation between these two parameters was considered to be due to the size dependent variation in grain alignment as $\lambda_{\rm max}$ is related to size of those grains which are aligned. It was found that the relation between $\lambda_{\rm max}$ and $R_{V}$ gives a meaningful estimation for extinction larger than ice threshold extinction ($A_{V}$ = 3.2) \citep{2001ApJ...547..872W}. Further a marginal correlation between $\lambda_{\rm max}$ and $A_{V}$ was found  \citep[e.g. in studies][]{2001ApJ...547..872W,2007ApJ...665..369A,2008ApJ...674..304W,2018MNRAS.475.5535I,2020ApJ...905..157V}. We have shown the dependence of $\lambda_{\rm max}$ on $A_{V}$ in Figure \ref{fig:lmax_av} for observed region NGC 2345. The members of cluster NGC 2345 are shown by red-filled triangles while other field stars are marked by black-filled circles and stars which are found to be intrinsically polarized from Section \ref{ser} are shown by black squares. The blue line represents the best fit straight line using the bootstrap sampling method as described in Section \ref{eff}. The best fit straight line shows the relation of $\lambda_{\rm max} = (0.09\pm0.05) \times A_{V} + (0.35\pm0.09)$. However, the linear relation between $\lambda_{\rm max}$ and $A_{V}$ is not statistically significant.

Figure \ref{fig:eff_lmax} shows the variation of $\lambda_{\rm max}$ with $P_{\rm max}/A_{V}$. It shows that with the increase in $P_{\rm max}/A_{V}$, $\lambda_{\rm max}$ decreases. A similar trend was also found by \citet{2001ApJ...547..872W,2007ApJ...665..369A,2012JQSRT.113.2334V,2016MNRAS.462.2343V,2018MNRAS.475.5535I}.

\section{Discussion} \label{diss}
A polarimetric study of a total of 197 sources of the cluster NGC 2345 and a nearby field region has been carried out. The polarization is found to be wavelength dependent for majority of stars with the average value of $\lambda_{\rm max} \approx 0.58 \mu m$ and $P_{\rm max} \approx 1.5\%$. No distinction in the distribution of degree of polarization was observed between cluster members and other observed stars. Below we discuss our results based on our findings.

\subsection{Orientation of dust grains towards the cluster}
The polarization vectors for the majority of stars in both cluster and field regions are found to be parallel to the GP. The parallelism of polarization vectors with the GP indicates that dust grains are aligned with the Galactic magnetic field in both regions. There are a few stars for which the polarization vectors are not parallel to the GP. We have also noticed the change in the polarization properties along with the increment of visual extinction near the distance of 1.2 kpc. Majority of stars with a larger deviation of $\theta_{V}$ from GP are found to lie before this distance (see Section \ref{heiles}). Also, while going away from the distance of 1.2 kpc, the values of $\theta_{V}$ are found to be similar to GP direction. Relatively smaller dispersion in $\theta_{V}$ for cluster members shows the better alignment of dust grains in the intra-cluster medium. It shows that there is a presence of different orientations of dust polarization (other than GP) for the foreground in the cluster region which causes a spread in position angles. It appears that grains beyond 1.2 kpc distance are aligned with the Galactic magnetic field. However, foreground dust may have a magnetic field significantly different from the GP. The magnetic field orientation in the foreground is not as organized as in the cluster region. Additionally, the decreasing trend of $\theta_{V}$ with distance beyond the distance of 1.2 kpc reveals the spatial variation of the magnetic field over different scales.
\subsection{Implications for grain alignment mechanisms}
The differential dust concentration in the cluster region is also observed. We have found a decreasing trend of polarizing efficiency with the radial distance of the cluster which shows the increased alignment efficiency towards the cluster centre. In previous studies of the cluster NGC 2345, it was found that the cluster is affected by differential reddening with the variation of $E(B-V)$ is from 0.5 to 1.2 mag with the average value of 0.66 $\pm$0.13 mag \citep[]{1974A&AS...16...33M,2015AJ....149...12C,2019A&A...631A.124A}. We have noticed that $P_{\rm max}$ in the core region of the cluster is higher than that in the coronal region, which is similar to that we found for the $E(B-V)$ values. In addition to this, no specific variation of $\theta_{V}$ with the radial distance is seen which implies that the orientation of the alignment factor is not much changing inside the cluster. The decreasing trend of $\lambda_{\rm max}$ towards the cluster centre shows that the average size of aligned grains reduces towards the centre of the cluster NGC 2345. The observed polarizing efficiency [$P_{\rm max}/E(B-V)$] for the majority of field stars is found to be more than the estimated value for the Galaxy, whereas the polarizing efficiency for cluster members is found to be similar or less than that of the estimated average value for the Galaxy \citep[see][]{2002ApJ...564..762F}. Also, the estimated average values of polarizing efficiency of the Galaxy of 3.8 $\%/mag$ is found to be in between to that of the average value for the field and member stars of  5.5 and 2.9 $\%/mag$, respectively. The lower polarizing efficiency of cluster members indicates the less polarizing efficiency of the intra-cluster medium. It seems that starlight from cluster members has been depolarized due to the nonuniform alignment of dust grains in the foreground and intra-cluster medium. The dispersion in $P_{max}$ for cluster members is compatible with the differential reddening in the cluster. Similar results were found by \citet[][]{2012MNRAS.419.2587E,2013ApJ...764..172P}.
Polarizing efficiency ($P_{V}/A_{V}$) is found to follow a power law relation with extinction, $A_{V}$ with a slope of -0.67. Differences in alignment of grains, grain properties (size), and magnetic fields can cause variation in $P_{V}/A_{V}$ and different slopes ($\alpha$) from region to region. There have been studies of different regions of ISM with similar power law behaviour \citep[e.g. in][]{1992ApJ...399..108G,1995ApJ...455L.171G,2008ApJ...674..304W,2014A&A...569L...1A,2015AJ....149...31J,2015MNRAS.448.1178H,2015ARA&A..53..501A,2021AJ....161..149S}, whereas for a few lines of sight, $\alpha$ was found relatively higher \citep[][]{1995ApJ...448..748G,2001ApJ...547..872W,2014ApJ...793..126C,2017ApJ...849..157W,2018MNRAS.475.5535I,2019ApJ...873...87M} showing lesser efficiency of grain alignment at higher extinction. According to \citet{1992ApJ...389..602J}, the slope of $\sim$ -0.5 for moderate extinction is expected due to the effect of magnetic field turbulence.

The slope of $\lambda_{\rm max}$ and $A_{V}$ relation in our study is found similar to that found by \citep[e.g.][]{2008ApJ...674..304W,2018MNRAS.475.5535I}. However, in some studies \citet[][]{2007ApJ...665..369A,2020ApJ...905..157V}, a relatively smaller slopes ($\approx$ 0.03) were found. According to radiative alignment theory, grains are aligned when anisotropic radiation is sufficiently strong to spin-up grains to superthermal rotation \citep{2021ApJ...908..218H}. Hence, with the increase in $A_{V}$, the aligning radiation becomes weaker and the gas density is higher, so that only larger grains with larger radiative torque efficiency can be aligned, corresponding to the increase in the size of the smallest aligned grains  \citep{2021ApJ...908..218H}. Therefore, the increase of $\lambda_{\rm max}$ with $A_{V}$ is expected. Additionally, we have found the decreasing trend of $\lambda_{\rm max}$ with the increase in $P_{\rm max}/A_{V}$ which shows with the increase in $P_{\rm max}/A_{V}$ (i.e. larger efficiency of alignment) the average size of aligned dust grains reduces. It suggests that smaller grains are also aligned with the larger alignment efficiency so the average size of aligned grains shifted towards a smaller value, while for lower alignment efficiency, relatively larger grains are aligned so the average size is large as a result $\lambda_{\rm max}$ show higher value. Our results shown in Figure \ref{fig:spatial} reveal an overall increase of $P_{\rm max}/A_{V}$ and decrease of $\lambda_{\rm max}$ toward the cluster center. This may be caused by the increase in the local radiation field due to stars within the cluster that increases the alignment efficiency of small grains, following the radiative torque alignment theory \citep{2021ApJ...908..218H}.

\section{Summary and Conclusions}
\label{summ}
Using the polarimetric observations of 197 stars in the cluster NGC 2345 and field regions, we have found a single distribution of degree of polarization and position angles. The majority of polarization vectors are found nearly parallel to the direction of Galactic parallel indicating the alignment of dust grains possibly with the Galactic magnetic field. An increment in the dust concentration seems to occur with $A_{V}$ of $\sim$ 1.15 mag near the distance of 1.2 kpc towards the line of sight which is also accompanied by the change in polarization properties. The variation of polarization with wavelength is fitted by Serkowski empirical relation, yielding the maximum value of polarization and its corresponding wavelength as 1.55\% and 0.58 $\mu m$, respectively. This indicates that the polarization in the direction is due to the foreground ISM and the average size distribution of dust grains is similar to the general ISM dust grains. The increment of wavelength for maximum polarization with extinction accompanied by its decrements with increasing polarizing efficiency could be a result of radiative torque alignment of grains. The increasing polarizing efficiency and decreasing wavelength of maximum polarization towards the cluster centre indicates about increasing local radiation field within the cluster which supports the radiative torque alignment theory.

\section*{Acknowledgement}
We  thank the referee for his/her useful comments and suggestion. Part of this work has made use of data from the European Space Agency (ESA) mission {\it Gaia} (\url{https://www.cosmos.esa.int/gaia}), processed by the {\it Gaia} Data Processing and Analysis Consortium (DPAC, \url{https://www.cosmos.esa.int/web/gaia/dpac/consortium}). Funding for the DPAC has been provided by national institutions, in particular the institutions participating in the {\it Gaia} Multilateral Agreement. T.H. acknowledges the support by the National Research Foundation of Korea (NRF) grants funded by the Korea government (MSIT) through the Mid-career Research Program (2019R1A2C1087045). S.S. acknowledges  Biman J Medhi for his support in initiating the project, Archana Soam for reading initial draft version of the paper, and Priyanka Srivastava for help during observations. 

\section*{Data availability}
The data underlying this article is available in the article.
\bibliographystyle{mnras}
\bibliography{ref}


\end{document}